\documentclass{aa}  
\usepackage{graphicx}
\usepackage{balance}
\usepackage{txfonts}
\usepackage{natbib}
\bibpunct{(}{)}{;}{a}{}{,} 

\def\lc#1{#1}
\def\pn{\pm\hphantom{1}}
\begin{document} 

   \authorrunning{P. Hadrava et al.}
   \titlerunning{Spectroscopy of UU~Cas}
   \title{Spectroscopy of the massive interacting binary UU~Cassiopeiae}

 \author{P. Hadrava \inst{1}\fnmsep\thanks{Orcid numbers: PH 0000-0002-4518-3918, MC 0000-0003-2050-1227,
 GD 0000-0001-9392-6678, SYuG 0000-0001-7380-1317, MIJ 0000-0002-8591-4295, DK 0000-0002-5859-7220,
 HM 0000-0002-4544-3070, REM 0000-0002-6245-0264, JP 0000-0001-8535-7807, SZ 0000-0003-2526-2683}
 \and M. Cabezas \inst{1,2}
 \and G. Djura\v{s}evi\'{c} \inst{3}
 \and J. Garc\'{e}s \inst{4} 
 \and S. Yu. Gorda \inst{5} 
 \and M. I. Jurkovic \inst{3}
 \and D. Kor\v{c}\'{a}kov\'{a} \inst{6} 
 \and\\ H. Markov \inst{7}
 \and R. E. Mennickent \inst{4}
 \and J. Petrovi\'{c} \inst{3} 
 \and I. Vince \inst{3}
 \and S. Zharikov \inst{8}
          }

 \institute{Astronomical Institute of the Academy of Sciences of the Czech Republic,
        Bo\v{c}n\'{\i} II 1401/1, 141 00 Praha 4, Czech Republic\\
        \email{had@asu.cas.cz}
 \and Institute of Theoretical Physics, Faculty of Mathematics and Physics, Charles University,
        V Hole\v{s}ovi\v{c}k\'{a}ch 2, 180 00 Praha 8, Czech Republic
 \and Astronomical Observatory Belgrade, Volgina 7, Serbia 
 \and Universidad de Concepci\'{o}n, Departamento de Astronom\'{\i}a, Casilla 160-C, Concepci\'{o}n, Chile
 \and Kourovka Astronomical Observatory, Ural Federal University, Mira str., 19, Yekaterinburg, 620002 Russia
 \and Astronomical Institute of Charles University,
        V Hole\v{s}ovi\v{c}k\'{a}ch 2, 180 00 Praha 8, Czech Republic
 \and Institute of Astronomy and NAO, Bulgarian Academy of Sciences, Bulgaria
 \and Universidad Nacional Aut\'{o}noma de M\'{e}xico, Instituto de Astronom\'{\i}a, AP 106, Ensenada 22860, BC, M\'{e}xico
    }

   \date{Received 28 October 2021; accepted 7 January 2022}

  \abstract
  {The eclipsing close binary UU~Cas is an interacting massive double-periodic system with a gainer star partly hidden in an accretion disk.}
  {In order to study  the physics of the accretion process in greater detail, along with the structure and dynamics of the circumstellar matter in the system, we supplement our previous results obtained from photometry with an analysis of the spectra of UU~Cas.}
  {We collected all available spectra used in previous publications on UU~Cas and we acquired new ones. The method of disentangling was applied to this set of spectra spanning the years 2008--2021. The orbital parameters were disentangled and a fit of the separated component spectra by synthetic ones has been used to determine the physical parameters of the component stars. We compared the results to models of the evolution of interacting binaries.}
   {We found that in addition to the dominant role of the donor star and a weak contribution of the gainer, the line profiles are strongly influenced by the circumstellar matter. The absorption lines reveal a presence of a disk wind emanating above the orbital plane. The variability of H$\alpha$ emission yields evidence of changes in the structure of the circumstellar matter on a timescale of several orbital periods.}
   {}


   \maketitle
   

%

\section{Introduction}

The eclipsing spectroscopic close binary star UU~Cas (BD +60 2629, $\alpha_{2000}$ = 23h 50m 39.52s, $\delta_{2000}$ = +60$^\circ$ $54'$ $39.14''$) consists of two interacting early B-type stars, with the more massive and hotter gainer almost hidden within the circumstellar disk.
This binary (with an   orbital period about 8.5 days) is a very complex system, thereby presenting a challenging and valuable laboratory for studying the behaviour of gas dynamics and the configuration of circumstellar and circumbinary envelopes. Its relatively high level of brightness ($V=9.74$) allows for necessary and sufficiently accurate  observed photometric and spectroscopic data to be obtained for the purpose of constructing detailed models.

 The basic properties of the system UU~Cas that were derived mainly from photometry have been updated and summarised by \citet{2020A&A...642A.211M}. 
In the current work, we present our previous and more recent spectroscopic observations, along with the results obtained from their analyses.

 The spectrum of UU~Cas appears, at first sight, to be a typical example of a single-lined spectroscopic binary (SB1) with numerous weak metallic lines and stronger helium and hydrogen lines moving in phase with the donor star resolved as the brighter component from the photometry.
The occurrence of an emission in the Balmer hydrogen lines testifies to the presence of circumstellar matter, most probably the disk identified in the photometric solution \citep{2020A&A...642A.211M},  
in which the gainer star is partly hidden.
It also indicates that the hydrogen lines are unsuitable for radial velocity (RV) measurements or for a determination of the orbital parameters.

 It has been noted by \citet{2010POBeo..90..159M,2011BlgAJ..15...87M} 
that the line He\,I 5875.6\,{\AA} exhibits a splitting into two absorption components around phases 0.69 and 0.12.
The stronger component was red-shifted at the phase of 0.12 and blue-shifted at 0.69.
\citet{2017AstBu..72..321G} 
confirmed this finding and presented plots from his spectra of line profiles of the same line at phases 0.31 and 0.76, which convincingly display a double-core structure (and possibly also a weak static emission, most probably of a circumstellar origin).
This implies that  in the spectra there is a non-negligible contribution of lines of the second component or the circumstellar matter, which may distort the results of interpretation of the spectra as an SB1. 
Using a fit of line profiles with three Gaussian curves, Gorda 
obtained RV curves of both components of this double-lined spectroscopic binary (SB2) and found a mass ratio for the system that indicated that the previously unseen component is about twice as massive than the visible one.
This is contrary to the previous assumptions, dating back to the first spectroscopic study by \citet{1934ApJ....79...84S}, 
which presumed that the invisible companion would be expected to have a smaller mass than the visible one.
The higher mass of the gainer is, however, also quantitatively consistent with the photometric solution and an assumption that the brighter star is close to its Roche lobe.
This change of view motivated also a reversal of designation of the primary and secondary star component in some recent literature, which is, however, confusing from the observational as well as evolutionary point of view.
Here, we thus denote the original `primary star', which is lighter, cooler but observationally brighter as the `donor'. The `secondary component', which is more massive, hotter but partly obscured by the disk, and, hence (despite its higher intrinsic luminosity), observationally fainter, is identified as the `gainer'.

 The photometric solution \citep{2020A&A...642A.211M} revealed a presence of a long period of about 270 days, as well as irregular variations in brightness on a short timescale caused most probably by changes in the UU~Cas circumstellar matter structure.
The presence of the accretion disk around the more massive and hotter gainer, which is almost hidden within it, and its complex structure changing over the time, may distort the measurements of radial velocities (RVs), not only in the Balmer lines.
To verify and further develop the present understanding of the system UU~Cas, we thus collected and analyzed all available sets of its spectra. 
In Section~\ref{SObs}, we list all these datasets.
The method and results of the analysis are described in Section~\ref{SAnal}.
The interpretation of the results is then discussed in Section~\ref{SDis}.
The conclusions are summarised and further prospects outlined in Section~\ref{concl}.


\section{Observations}\label{SObs}

In addition to the RVs published in the literature \citep{1934ApJ....79...84S}, 
six sets of spectra are used in this study, collected by the Rozhen National Astronomical Observatory (RNAO) from 2008 to 2014, at San Pedro de Martir Observatorio Astronomico Nacional (OAN) in 2011, at Gothard Astrophysical Observatory (GAO) in 2012, at Apache Point Observatory (APO) in 2012 and 2013, at Kourovka Astronomical Observatory (KAO) in 2017 and 2018, and at Ond\v{r}ejov Observatory of the Astronomical Institute (AIO) in 2020 and 2021 (cf. Table~\ref{datasets}).

 The RNAO spectra were observed using the {Coudé} spectrograph with CCD detector attached to the 2m telescope.
The observations were carried out in five different spectral regions (about 200\,{\AA} wide) centered on wavelengths of 4500\,{\AA} (near the Mg\,II 4481 line), 4700\,{\AA} (near the He\,I 4713 line), 4800\,{\AA} (near the H$\beta$ line), 5800\,{\AA} (near the He\,I 5875 line) and 6500\,{\AA} (near the H$\alpha$ line).
Some of these spectral regions, besides Mg\,II line and the hydrogen and helium lines, contain also metallic ion lines (N\,II, Si\,III, O\,II, Fe\,III) suitable for RV measurements and line profile analyses.
The {Coudé} spectrograph resolving power, the signal-to-noise ratio (S/N) and spectral dispersion were about 30000, 50--150 and 0.2\,{\AA}/pixel, respectively.
Due to the detector resolution, the actual resolving power of the observed spectra of UU~Cas is reduced to about 15000.
The observed data were processed in the usual way: bias and dark frames were subtracted and flat-fielded, with the cosmic particles filtered out. 
Wavelength calibrations were done using calibration spectra of a Th-Ar calibration lamp.
The one-dimensional (1D) spectra were obtained by averaging the weighted measured intensities in the direction perpendicular to the dispersion.
These were normalised to the continuum approximated by spline functions.
The processing was performed in the IRAF Package.\footnote{\texttt{IRAF} is distributed by the~National Optical Astronomy Observatories, operated by the~Association of Universities for Research in Astronomy, Inc., under contract to the~National Science Foundation of the~United States.
}

 Beside the RNAO observations, we used 30 spectra obtained with the Echelle spectrograph at 2.12m telescope of OAN on ten consecutive nights in 2011.
This set of spectra thus uniformly covers one period and provides an `instantaneous' picture of the system that facilitates the process of distinguishing the orbital changes from the long-term variations.
In the processing of the observed spectra, the bias frames were subtracted, cosmic particles filtered out, orders extracted, the wavelength calibration was done using calibration spectra of a calibration lamp, and the orders were combined.
The spectrograms cover wavelengths from about 3800\,{\AA} to 7600\,{\AA}. 
The maximum achievable resolving power at 5000\,{\AA} is about 18000.
Unfortunately, the S/N is highly variable, depending on the wavelength, and {decreases} rapidly near the shortest and longest end of the observed spectral region.
We thus used only the spectra in the region 4300\,{\AA} to 7100\,{\AA} . 
The three spectra from the last night suffer from a particularly high degree of noise and were thus excluded or diminished by a low weight in our calculations.
An internal random error of the wavelength measurements from the measured variation of the difference of wavelengths between the interstellar Na D1 and D2 lines was estimated.
These measurements were carried out using the {\sc Spefo} program package written by Ji\v{r}\'{\i} Horn \citep[cf.][]{1996ASPC..101..187S} 
and the rms error obtained from 30 spectra was 0.8\,km/s.
An additional correction of RVs was found via a disentangling process in the same spectral region.

\begin{table}
\caption{Sets of spectra analyzed in this paper}\label{datasets} 
\begin{tabular}{lccccc}
Obs.  & instrument &  $R$  & $N$ & S/N & dates \\ \hline
RNAO  & 2m / {Coudé}  &15000 & 49 & 95 &4811--6964\\ 
OAN   &2.12m / Echelle&18000 & 30 & 68 &5580--5588\\
GAO   &0.5m / Echelle &18000 &  8 & 54 &6203--6206\\
APO   & 3.5m / ARCES  &30000 &  3 & 53 &6227,6616\\
KAO   &1.21m / Echelle&15000 & 27 & 41 &7779--8221\\
AIO   & 2m / OES      &50000 & 17 & 35 &8925--9294\\
\end{tabular}
\tablefoot{Obs. denotes the observatory (cf. the text), instrument denotes the telescope or spectrograph, $R$ the resolving power, $N$ the number of exposures, S/N is the signal-to-noise ratio, and dates are given as JD$-$2450000.
}
\end{table}

 Eight CCD spectra were taken at Gothard Observatory of E\"{o}tv\"{o}s Lor\'{a}nd University with the Echelle spectrograph 
attached to the RC telescope with primary mirror of 50 cm in diameter \citep{2014CoSka..43..183C} 
over the period of \lc{2--5 October} 2012.
The observed spectra cover wavelength interval from 4400\,{\AA} to 7000\,{\AA} with about 30000 measured wavelength positions. Wavelength calibrations were done using the observed spectra of standard star $\beta$ Oph.
The observed spectra were processed using the IRAF Package. The maximum achievable spectral resolving power at 5000\,{\AA} was about 18000.
The S/N {for the GAO spectra is also highly wavelength dependent and it drops down to values between 5 and 15 towards the short-wavelength end of the observed range.
Fortunately, in the} vicinity of the He\,I 5875\,{\AA} line, the S/N (measured in wavelength interval from 5850\,{\AA} to 5865\,{\AA}) is much higher (from about 50 to 75, except for observation at HJD 2456203.4480 for which it {is about 20 only}).

 Two consecutive exposures, at 15 minutes each, were taken on one night in 2012 and one 20 minute exposure in 2013 at APO with ARCES.
This echelle spectrograph, with a resolving power of approximately 30000, provides complete spectral coverage from 3200\,{\AA} to 10000\,{\AA}.

 Twenty seven spectra were obtained using the fiber-fed echelle spectrograph attached to the 1.21m telescope at Kourovka Observatory of Ural Federal University \citep[for details cf.][]{2017AstBu..72..321G}. 
Some of them could not be used in all orders.

 Finally, seventeen spectra of UU~Cas together with the necessary dark frames, flat-fields, and Th-Ar comparison spectra were acquired using the Ond\v{r}ejov Echelle Spectrograph (OES, resolving power 50000) attached to the {Coudé} focus of 2m Perek's Telescope at Ond\v{r}ejov Observatory of the Astronomical Institute of Czech Academy of Sciences \citep[cf.][]{2004PAICz..92...37K,2020PASP..132c5002K}. 
These spectra were processed using a semi-automated IRAF Package prepared by M. Cabezas for this purpose, in which the standard procedures for echelle spectroscopy were used, including flat and bias correction, cosmic rays removal via a Laplacian algorithm \citep{2001PASP..113.1420V}, 
wavelength calibration using Th-Ar lamp, order merging, the rectification of spectra, and heliocentric calibration.


\section{Analysis of the UU~Cas spectra}\label{SAnal}

Regarding the variability of the system UU~Cas on different timescales found in our previous study of photometry \citep{2020A&A...642A.211M} 
along with the need to reveal a nature of long-term changes, we collected all available spectroscopic data and decided to analyze them first separately for each observational run.
 For our analysis, we used the {\sc Korel} code for the Fourier disentangling \citep[cf. ][]{1995A&A..114..393H}. 
We fixed the orbital period to the value of 8.519296 days found in the previous photometric study. 
We solved for eccentric orbits in several regions of spectra and we found various values of the eccentricity up to about 0.03 as well as very different values for the periastron longitude. 
This implies that the eccentricity, if there is any, is not measurable in the current data. 
We thus fixed the eccentricity to zero, which is often true for short-period binaries owing to the tidal circularisation of their orbits.
The longitude of periastron has been fixed equal to 90$^\circ,$ so that the time of periastron passage corresponds to the primary minimum, namely, the eclipse of the donor.
Its initial value was set as JD=2443132.25696, according to the photometric ephemeris \citep{2020A&A...642A.211M}; 
it was, however, converged to get the best fit for each set of spectra.

\subsection{Disentangling of He lines}
To facilitate a comparison with previous results, in particular with those obtained by \citet{2017AstBu..72..321G}, 
we focus first on the spectral region 5860--5884\,\AA,{} which is dominated by the He\,I line 5875.6\,\AA{}.
We sampled this region in 4096 bins with a step of 0.3\,km/s per bin.
When we disentangled the OAN spectra in the standard way as an SB2 (cf. Solution 1 in Table~\ref{Tab0}), we got the profile of the He line of the donor star, which moves on a circular orbit with a semiamplitude. \lc{This agrees} well with the values of $K_d=191.2\pm2.7$ and $190.3\pm2.3$\,km/s found by Gorda for N\,II and He\,I lines, respectively. 
Using a synthetic spectrum as a template, we found a systemic velocity of the donor $\gamma_{d}=-71.9$\,km/s 
(while Gorda gave $-62.8\pm2.0$ and $-60.7\pm1.8$\,km/s, resp.).
We found the best fit of the line profile for rotational broadening of 115.1\,km/s. 
This is consistent with the value of 119\,km/s that corresponds to a synchronous rotation of the donor star if, according to the photometric solution \citep{2020A&A...642A.211M}, 
we accept its radius to be 20.8 $R_\odot$ with an inclination of 74.5$^\circ$.

 However, concerning the second component of the helium line, the SB2-disentangling of OAN spectra yielded results that are completely different from those presented by \citet{2017AstBu..72..321G}. 
The RV semiamplitude of this second component converged to zero (i.e. the mass ratio defined as $Q\equiv M_g/M_d=K_d/K_g$ diverged over 1800), while \citet{2017AstBu..72..321G} 
found a value $K_g=102.2\pm2.8$\,km/s.
In Solution 1, we started the convergence of $Q$ from a value of 2, which roughly corresponds to the estimate from Gorda.
As $Q$ diverges to infinity, the RV-semiamplitude of the second component approaches zero, but it cannot reach negative values.
If, however, we started from a negative value of $Q$, we arrived at Solution 2, which gave an even better fit of the observed spectra with practically the same profiles of the components, but the second component moving with a semiamplitude over 8\,km/s in phase -- instead of in antiphase with the donor star.

 A fit of the second component by a synthetic template showed a systemic velocity of $-156.0$\,km/s 
and a line width corresponding to a rotational broadening of 190.1\,km/s. 
It means that this component of the line is blue-shifted for about 84\,km/s with respect to the donor star, while Gorda gave a systemic velocity of the {gainer of} $-36.2\pm2.9$\,km/s, namely, it is red-shifted for 24.5\,km/s with respect to the donor. 
It is thus obvious that, at least in our spectra from season 2011, this second component of the helium line does not originate purely from the photosphere of the gainer star, but mainly from its extended envelope, and most probably from its parts around the centre of mass \lc{of the binary system}, which does not follow the orbital motion of the gainer.
The blue shift indicates an expansion of the absorbing gas.

\begin{figure} 
\setlength{\unitlength}{1mm}
\begin{picture}(90,120)
\put(0,0){\includegraphics[width=\hsize]{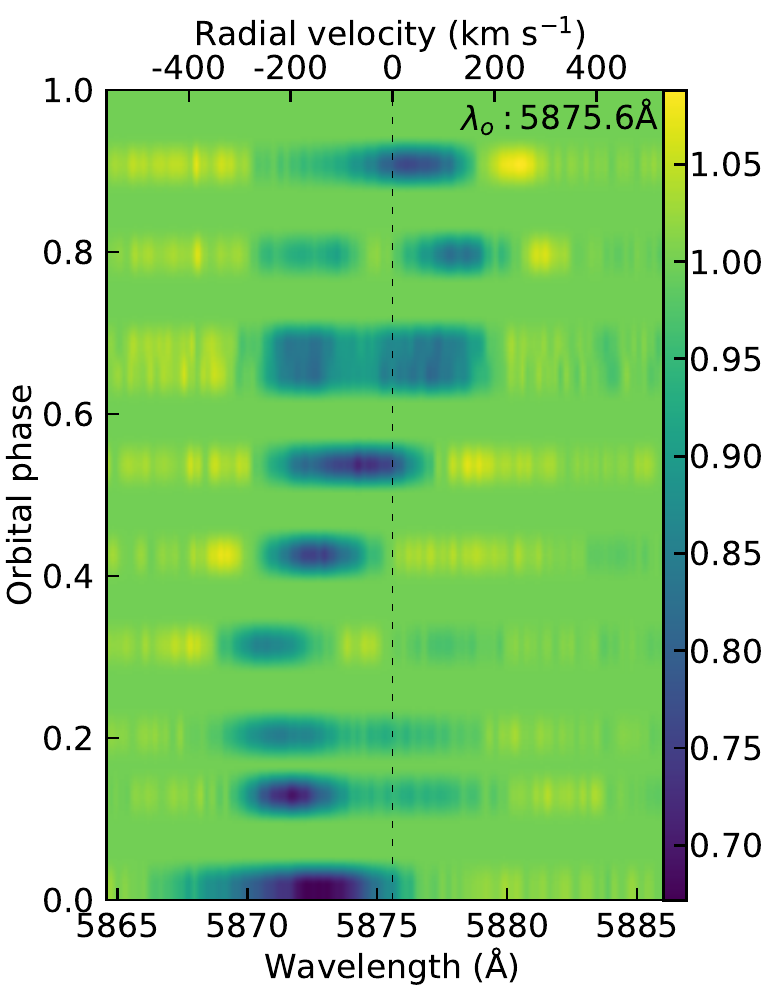}}
\end{picture}
\caption{ Dynamical spectra of the line  He\,I 5875.6\,{\AA} in OAN exposures
}
\label{P5860tr}
\end{figure}

 The \lc{line-strength factors} ($s$-factors, i.e. the natural logarithms of relative depths of a line) of this second component of the He line revealed that it is particularly strong in the vicinity of the primary minimum and \lc{it is a bit less enhanced also} around the secondary minimum.
This can also be seen in the dynamical spectra (cf. Fig.~\ref{P5860tr}), where the S-wave of the primary component appears to be enhanced around the conjunctions.
However, in the same figure, we can also recognise, in some phases around the quadratures, a weak S-wave (with a smaller amplitude) that corresponds to the motion of the secondary component star, that is, the gainer.

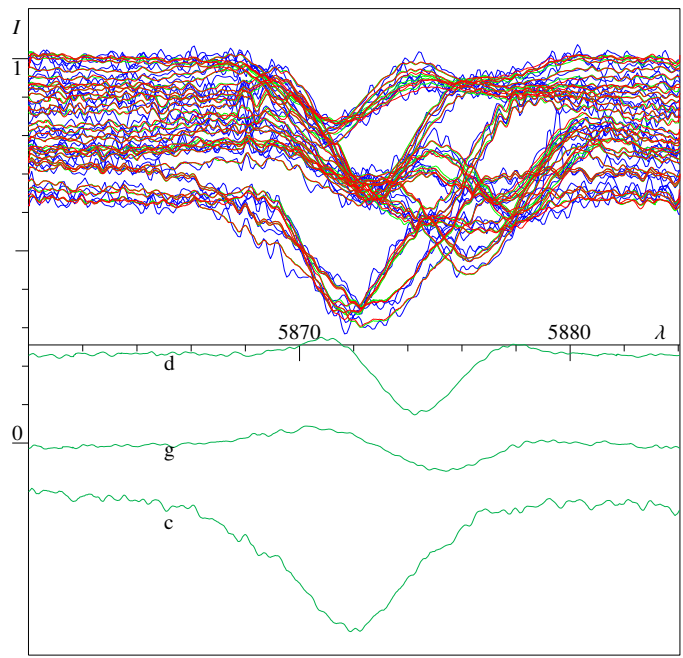
\begin{figure} 
\setlength{\unitlength}{1mm}
\begin{picture}(150,90)
\put(0,0){\includegraphics[width=\hsize]{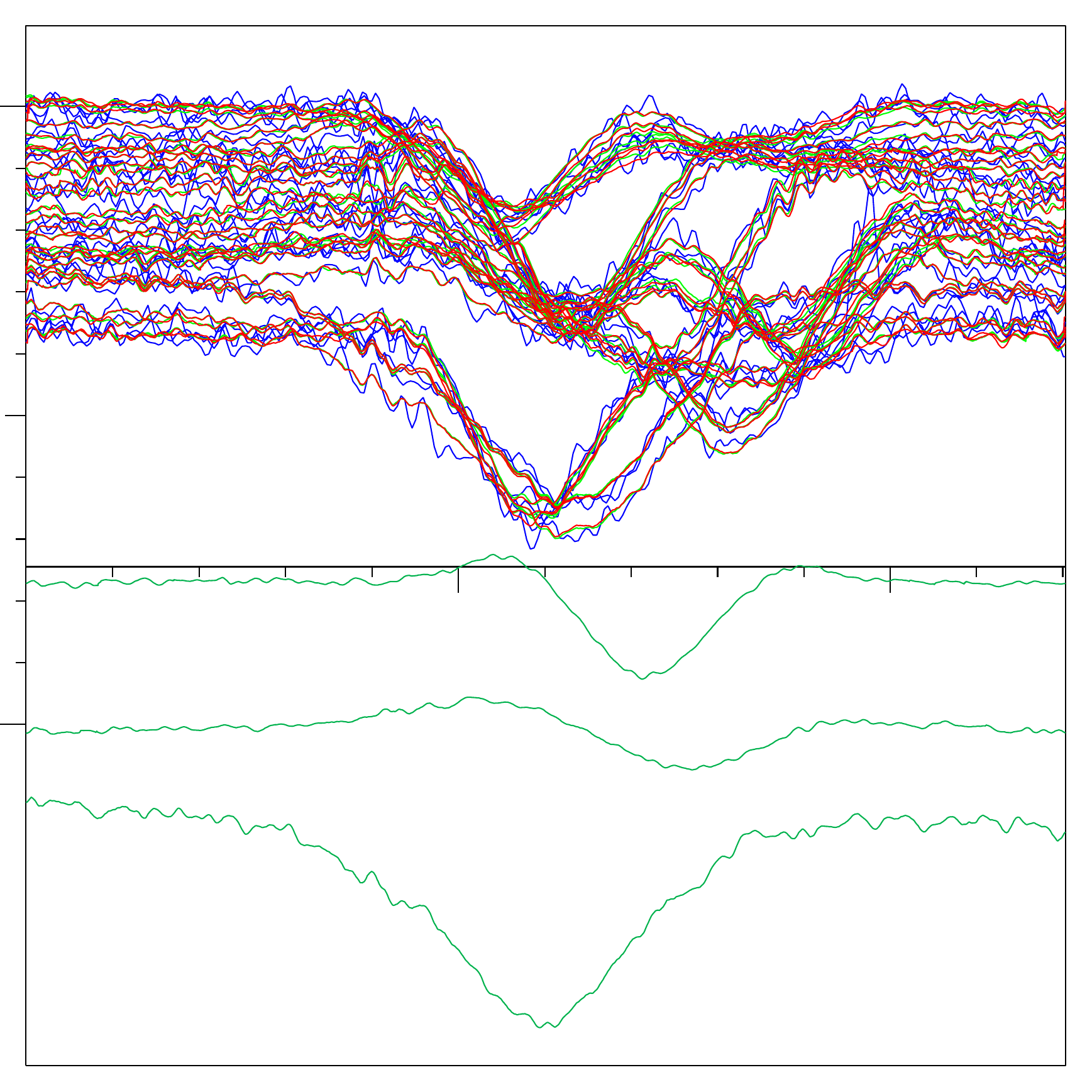}}
\put(0,84.5){\tiny $I$}
\put(0,79){\tiny 1}
\put(0,30.7){\tiny 0}
\put(35,44){\tiny 5870}
\put(70.5,44){\tiny 5880}
\put(84.5,44){\tiny $\lambda$}
\put(20,40){\tiny d}
\put(20,28.5){\tiny g}
\put(20,18.8){\tiny c}
\end{picture}
\caption{Disentangling of the region of He\,I line 5875\,\AA{} in OAN spectra of UU~Cas (Solution 3, cf. the text).
{Here and in the following, the intensity, $I,$ is given in units of the continuum level and the wavelength $\lambda$ 
is in {\AA}}}
\label{UUCas5875}       
\end{figure}

 This suggests that it is necessary to disentangle three components in the spectra -- the primary and secondary component that follow the orbital motion of the two stars, {namely, the donor and the gainer, respectively,} and a third component originating in the circumstellar matter.
The RVs of this third component can be solved for as free parameters.
However, our experience from many solutions showed that the RVs have a random scatter of several km/s and no regularity depending on the orbital phases.
We thus fixed the RVs of the third circumstellar component to zero.
The orbital parameters obtained from this three-component disentangling are given in Solution 3 in Table~\ref{Tab0} 
and the disentangled spectra can be seen in Fig.~\ref{UUCas5875}.
The blue lines in the upper part of this figure from the top downward are the observed spectra in chronological order.
The three dark green lines in the lower part denoted by letters `d', `g', and `c' are the disentangled line profiles of the donor, the gainer, and the circumstellar matter, respectively.
It is worth noting that there appear to be some slight emission wings of the helium line of the donor, indicating the non-negligible structure of its atmosphere. 
These emission wings were also visible in Solutions 1 and 2.
The observed (blue) profiles are overploted by light green lines reconstructed as a superposition of the disentangled profiles Doppler-shifted according to the disentangled orbital parameters and by red lines shifted according the RVs optimised for each exposure independently.
The great depth of the third circumstellar component around the primary minimum and its significant blue shift where both component stars have nearly the systemic RV explains why in the SB2-disentangling, it is preferable \lc{for the second profile to fit} to this circumstellar component, instead of the weak lines of the gainer.

 The rotational broadening of the donor found using the template-constrained disentangling is 111.5\,km/s and Doppler shift of \lc{$-66.7$\,km/s}.
For the gainer, we get a broadening of 158.0\,km/s and a Doppler shift of \lc{$-54.3$\,km/s, namely, slightly less} but also red-shifted with respect to the systemic velocity found from the donor, in agreement with Gorda's result.
The third component of the line profile corresponds to a broadening about 182.6 km/s and blue shift of \lc{$-226.6$\,km/s}.

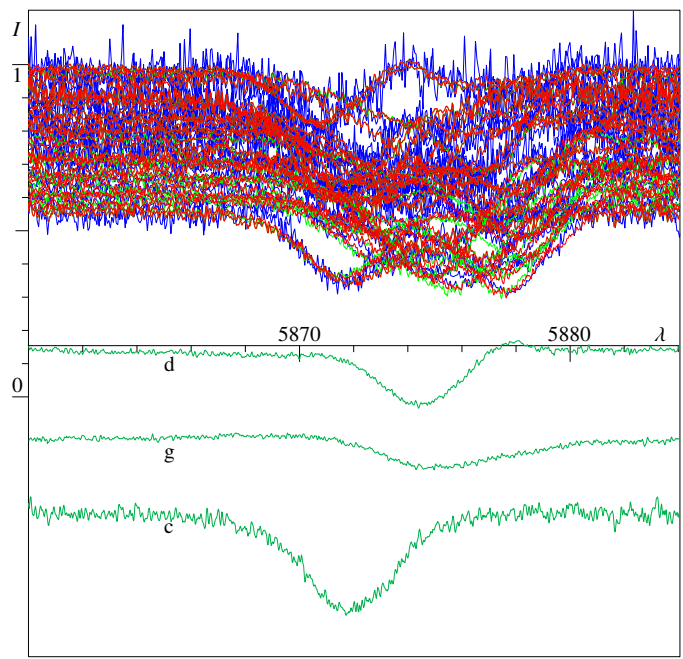
\begin{figure} 
\setlength{\unitlength}{1mm}
\begin{picture}(150,90)
\put(0,0){\includegraphics[width=\hsize]{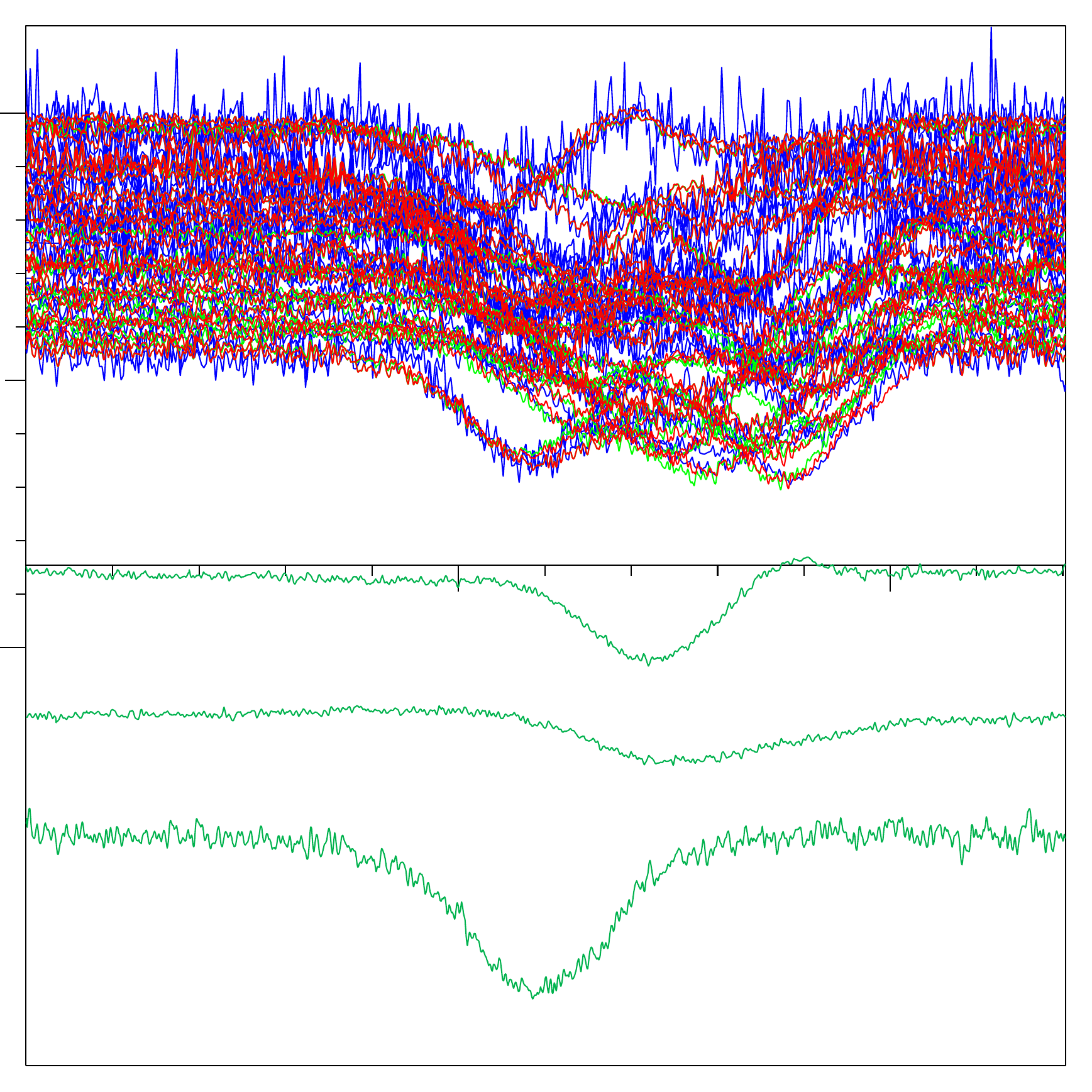}}
\put(0,84.5){\tiny $I$}
\put(0,78.3){\tiny 1}
\put(0,37.2){\tiny 0}
\put(35,44){\tiny 5870}
\put(70.5,44){\tiny 5880}
\put(84.5,44){\tiny $\lambda$}
\put(20,40){\tiny d}
\put(20,28.5){\tiny g}
\put(20,18.5){\tiny c}
\end{picture}
\caption{Disentangling of the region of He\,I line 5875\,\AA{} in KAO spectra of UU~Cas}
\label{K5860}   
\end{figure}

 When we applied the same procedure to the KAO spectra, we arrived at similar results.
The SB2-disentangling of all spectra separated the profile of the donor and the circumstellar component.
When only the spectra around the quadratures were taken into account, the second disentangled component switched over to the gainer.
Using the three-component disentangling, we separated from the donor both the gainer and the circumstellar line profile.
It is, however, interesting to note that while the spectra from the season 2017 were fitted quite precisely, the fit of spectra from 2018 was less precise and also the RVs had larger errors with respect to the disentangled orbital parameters.
This can be seen in Fig.~\ref{K5860}, where the green lines are well overlaid by the red lines for the exposures from 2017 at the top, but the red lines are displaced for the bottom exposures from 2018.
We thus disentangled the season of 2017 separately.
The orbital parameters disentangled from the spectra of 2017 are given in Solution 4.
Its higher precision can also be documented by smaller rms errors, namely, of 0.3 km/s and 0.2 km/s for $RV_d$ and $RV_g$ as compared to the values 10.1 km/s and 3.4 km/s for the solution of both seasons 2017 and 2018 altogether.
This change in \lc{behaviour of UU~Cas} occurred over the course of about ten orbital periods between the two KAO exposures, shown separated in Table~\ref{Tab1} by a horizontal line.

 The remaining eight KAO spectra from 2018 do not contain any exposure close to conjunction, which would define the profile of the circumstellar component in a three-component disentangling.
The SB2-disentangling of these spectra gives an unrealistic value of the mass ratio.

 Applying the same procedure to the RNAO spectra, we find Solution 5 in Table~\ref{Tab0} and disentangled line profiles that are similar to the previous ones. 
For the AIO spectra, we arrived at Solution 6, shown in Table~\ref{Tab0}, and the disentangled line profiles of the donor and gainer also similar to those obtained from the OAN spectra, while the line profile of the circumstellar component is more asymmetric with an emission in the long-wavelength wing. 

 The eight GAO exposures were taken at three different nights, so they  effectively represent only three points on the RV-curve and three spectra measured at different phases.
They are thus insufficient for an independent disentangling the three profiles of components and orbital parameters. 
This insufficiency is even more valid for the APO exposures, which represent two points only. 
Regarding the fact that the APO spectra were taken shortly after the GAO spectra, it is obvious that  they ought to be joined into one set.
There is no exposure close to the primary minimum where the circumstellar component is well pronounced, hence, it could not be reliably separated.
We thus fixed its profile using a template taken from Solution 3 of the OAN spectra and we calculated only its contribution to the overall profile by means of the $s$-factors.
The separation of the line profile of gainer was unstable, so we constrained it also by the template from OAN.
The resulting Solution 7 is given in Table~\ref{Tab0}. 

\begin{figure} 
\setlength{\unitlength}{1mm}
\begin{picture}(150,50)
\put(0,-3){\includegraphics[width=\hsize]{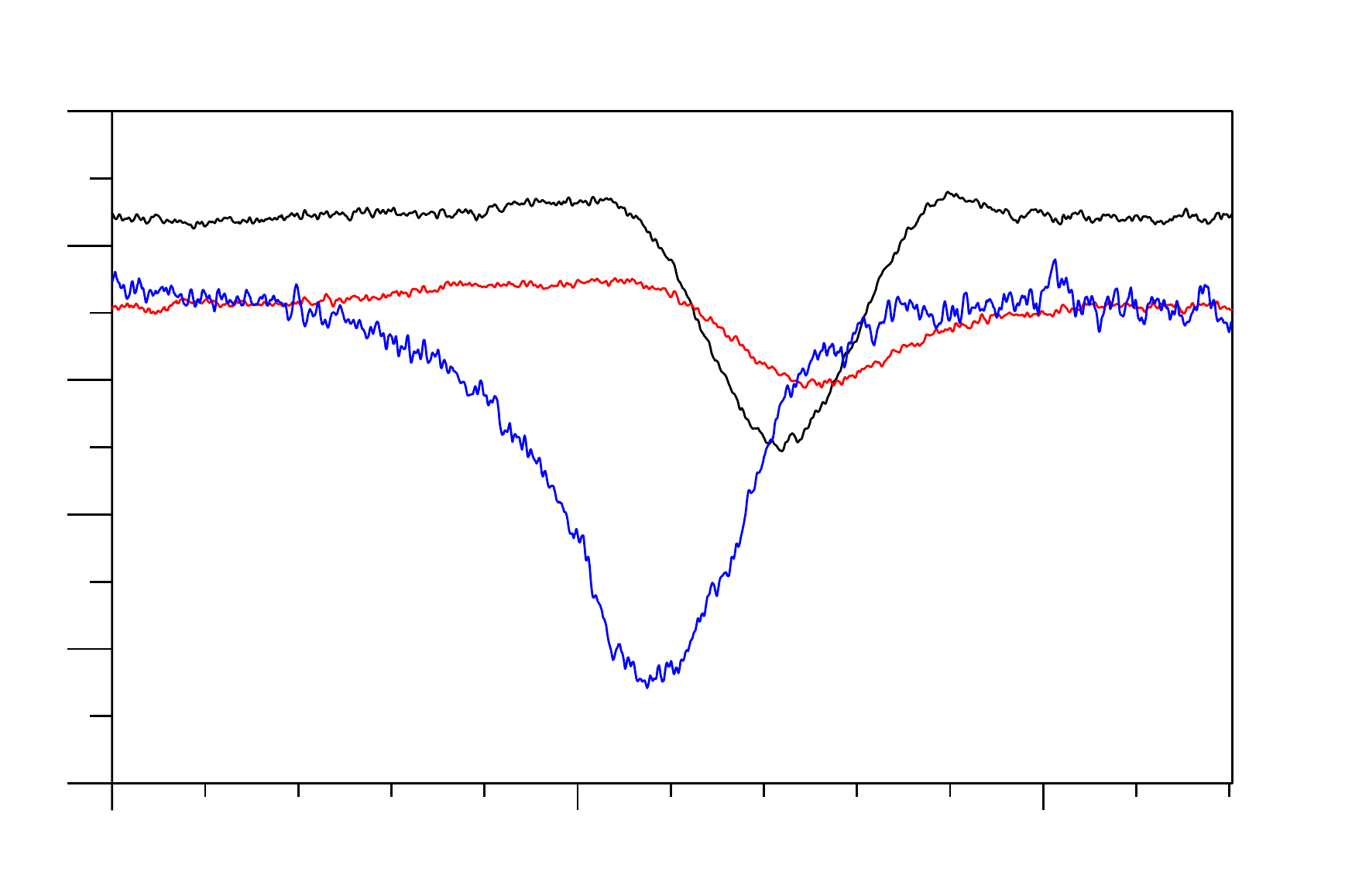}}
\put(2,46){\small $I$}
\put(0,40){\small 1.0}
\put(0,22){\small 0.8}
\put(0,4){\small 0.6}
\put(5,0){\small 5860}
\put(36,0){\small 5870}
\put(67,0){\small 5880}
\put(79,0){\small $\lambda$}
\end{picture}
\caption{Disentangled profiles of He\,I line 5875\,\AA{} of UU~Cas from all datasets (Solution 8). 
The depths of lines of donor (black) and gainer (red) correspond to their mean values during the orbit, while the depth of the circumstellar component (blue) corresponds to its maximum}
\label{spe}
\end{figure}
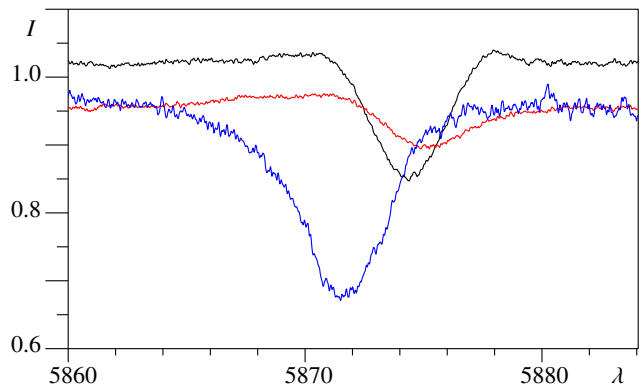

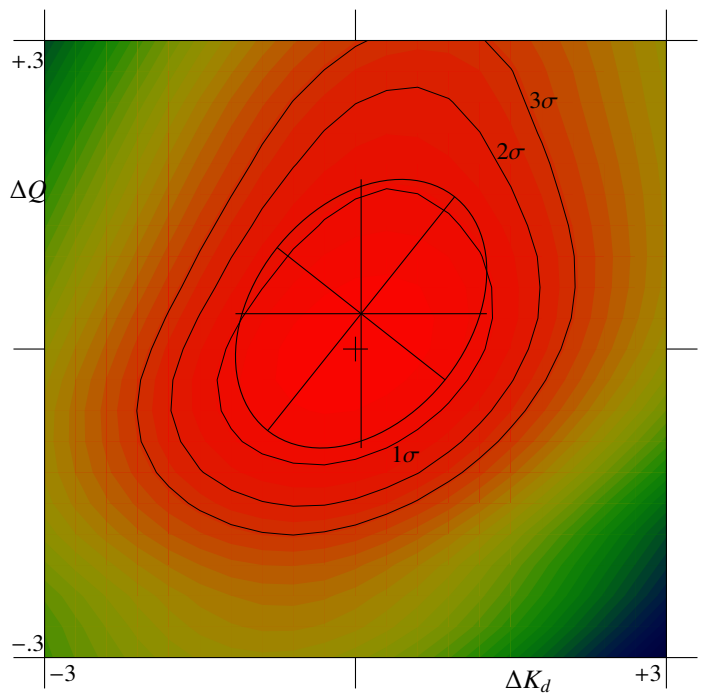
\begin{figure} 
\setlength{\unitlength}{1mm}
\begin{picture}(150,92)
\put(0.3,0){\includegraphics[width=\hsize]{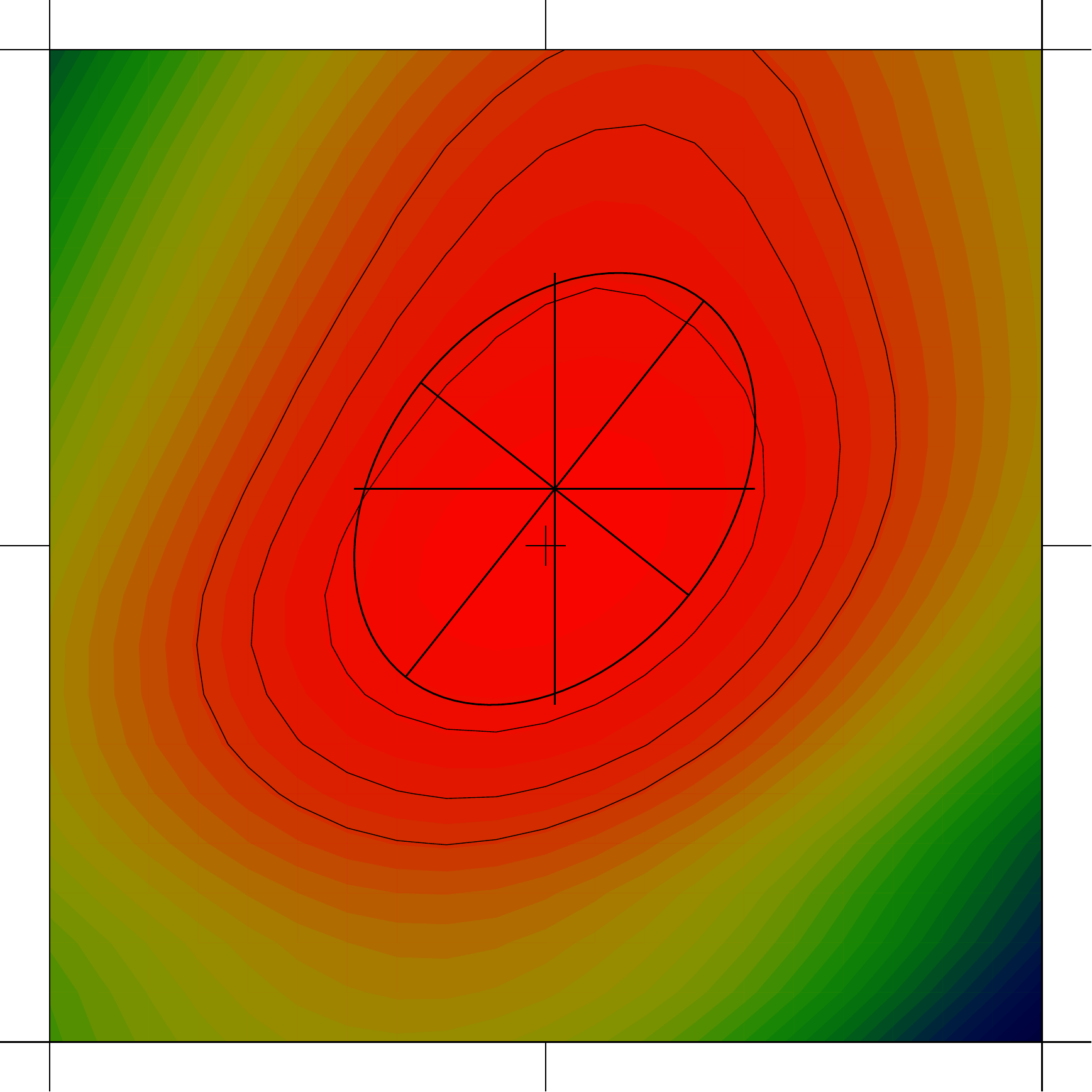}}
\put(65,0){$\Delta K_d$}
\put(5,1){\small $-3$}\put(82,1){\small +3}
\put(-0.2,65){$\Delta Q$}
\put(0,5){\small $-.3$}\put(0,82.3){\small +.3}
\put(50,30){\small 1$\sigma$}
\put(63.9,70.2){\small 2$\sigma$}
\put(68.2,77){\small 3$\sigma$}
\end{picture}
\caption{Probability distribution of Solution 8 in the cross-section of parameters $K_d$ (in km\,s$^{-1}$) and $Q$. 
The small cross in the centre of the figure corresponds to the highest probability {which is in the values given
in Table~\ref{Tab0}, the axes are labeled in differences from these values. The} centre of the probability ellipse 
obtained from the moments of the distribution is shifted to a higher value of $Q$. 
The three isolines give limits of 1, 2, and 3-$\sigma$ probability, which admits even higher values of $Q$}
\label{isolin}
\end{figure}

 The results from individual seasons showed only small changes in the disentangled profiles of the He line and the orbital parameters. 
It is thus worthwhile to also find  a mean solution from all available data.
If we join all sets together, we get 91 spectra with a mean S/N of about 40.
Their disentangling yields Solution 8 in Table~\ref{Tab0}.
The disentangled profiles are shown in Fig.~\ref{spe}. 
We also calculated the distribution of Bayesian probability of the disentangled orbital parameters
\citep[cf.][]{2016ASSL..439..113H}.
The cross-section in the subspace $\{K_d, Q\},$ \lc{shown in} Fig.~\ref{isolin}, reveals an asymmetry in $Q$ that indicates a possibility of even higher values of $Q$.
It should be noted that the errors of orbital parameters given in Table~\ref{Tab0} for the individual solutions are computed with \lc{line-strength factors} fixed to the values of the best found solution.
If we allow a convergence of the $s$-factors, then a wider region of the orbital parameters can fit the observed spectra comparably well and the actual errors of the orbital parameters may thus be higher.

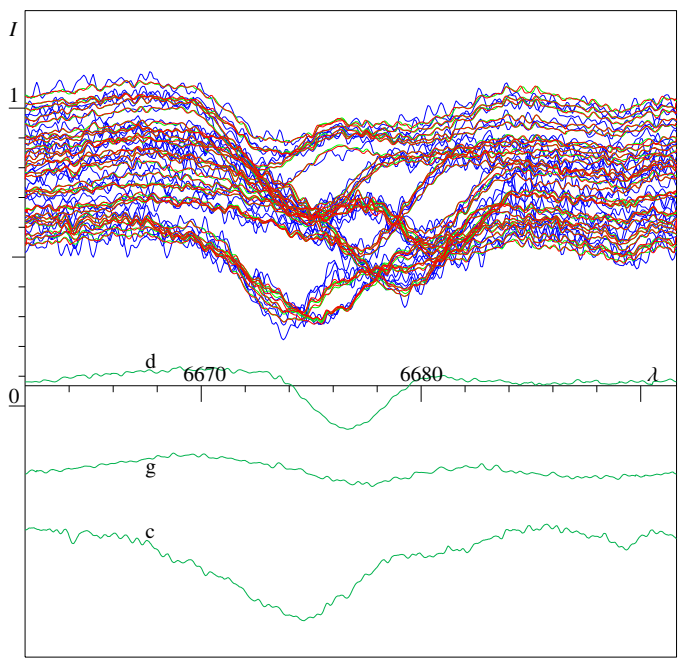
\begin{figure} 
\setlength{\unitlength}{1mm}
\begin{picture}(150,90)
\put(0,0){\includegraphics[width=\hsize]{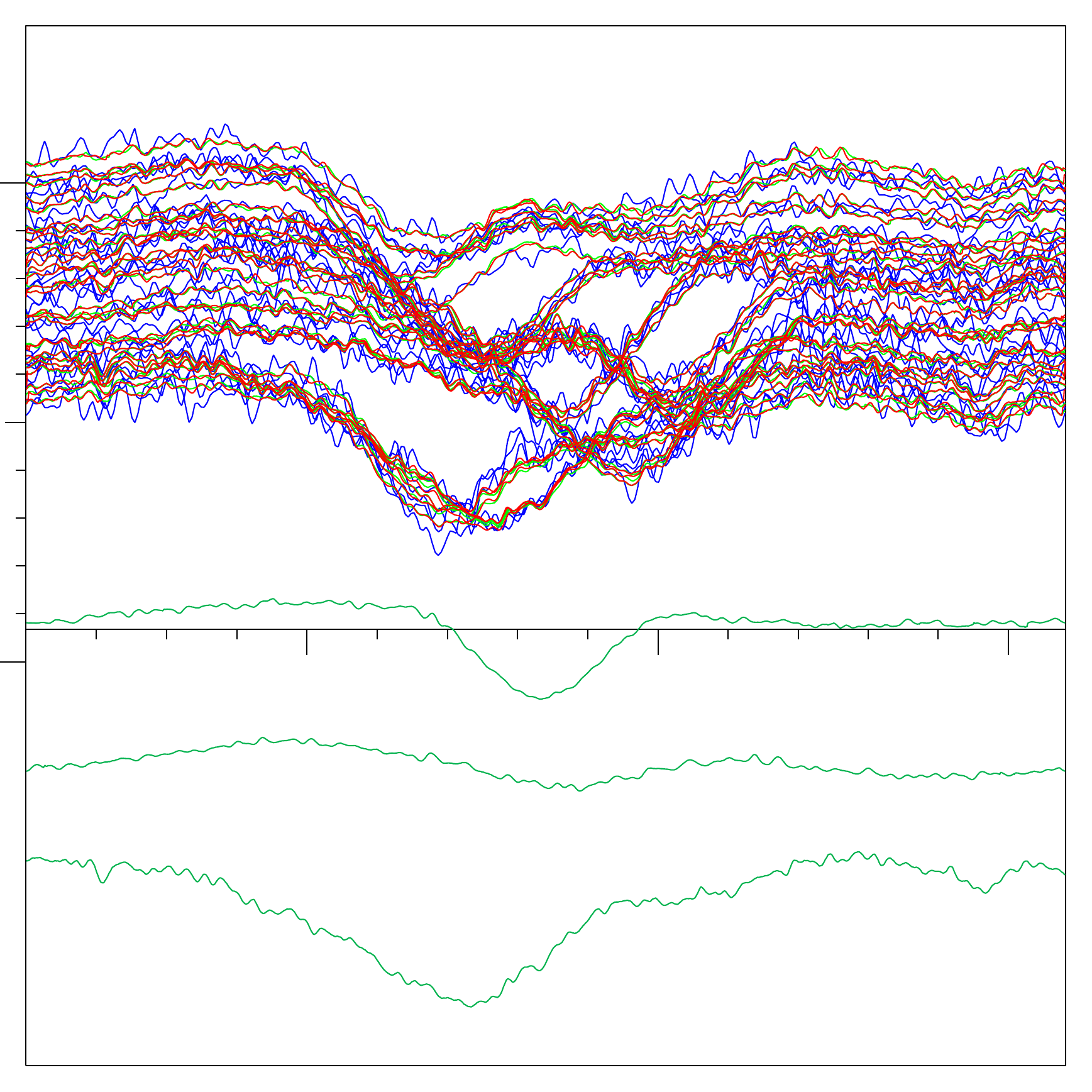}}
\put(0,84.5){\tiny $I$}
\put(0,75.7){\tiny 1}
\put(0,35.9){\tiny 0}
\put(23,38.8){\tiny 6670}
\put(51.5,38.8){\tiny 6680}
\put(84,38.8){\tiny $\lambda$}
\put(18,40.5){\tiny d}
\put(18,26.5){\tiny g}
\put(18,18){\tiny c}
\end{picture}
\caption{Disentangled profiles of He\,I line 6678\,\AA{} of UU~Cas from OAN spectra (Solution 9)}
\label{P6662}
\end{figure}

 Similar results were also obtained from disentangling of the He\,I line 6678\,{\AA}.
We sampled its vicinity in 2048 bins with a step of 0.65\,km\,s$^{-1}$.
The results from OAN spectra can be seen in Solution 9 and Fig.~\ref{P6662}.
The AIO spectra give Solution 10 for this line, however, the disentangled profiles have less apparent emission wings and a common solution of both these sets is worse than the separated solutions.
Only four RNAO spectra include this line, whereas the APO and some KAO spectra could not be used, so that there are altogether 80 spectra used for the overall Solution 11.

 The OAN, KAO, and AIO spectra enabled us to disentangle also the He\,I line 7065.2\,{\AA}.
The region 7054--7078\,{\AA} (which we have sampled in 2048 {bins} with a step of 0.5\,km\,s$^{-1}$) is contaminated by telluric lines.
We thus separated them as a fourth component \citep[cf.][]{2006A&A...448.1149H}.
The obtained orbital parameters are given in the Solution 12 in Table~\ref{Tab0}.

\begin{figure} 
\setlength{\unitlength}{1mm}
\begin{picture}(150,54)
\put(0,0){\includegraphics[width=\hsize]{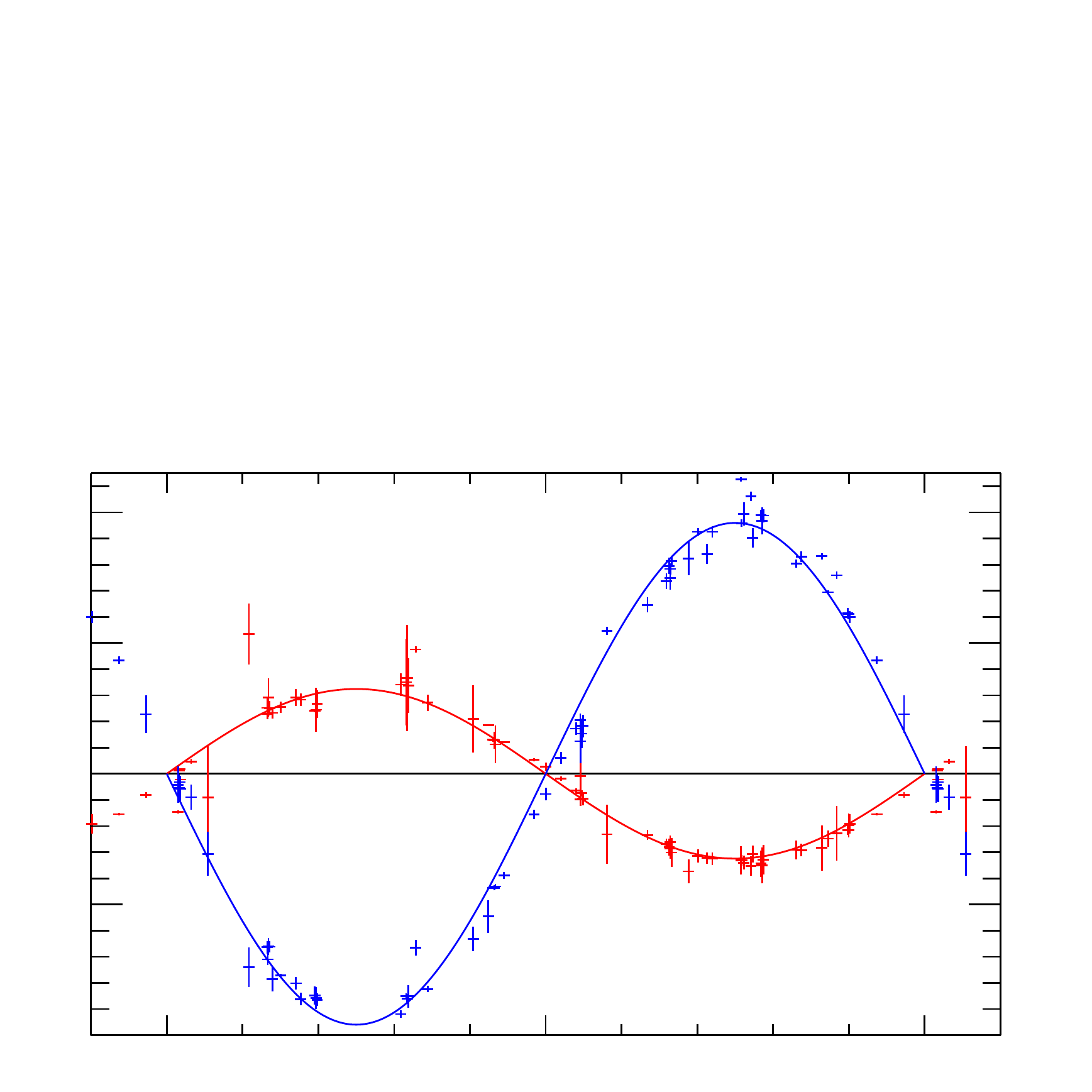}}
\put(0,46.6){\small +200}
\put(4,25.5){\small $0$}
\put(0,4.4){\small $-200$}
\put(13,1){\small 0}
\put(43,1){\small 0.5}
\put(76,1){\small 1}
\put(67,0){\small $\varphi$}
\put(2,40){\small $RV$}
\end{picture}
\caption{ {Phase dependence of mean} RVs in km\,s$^{-1}$ of donor (blue {line}) and gainer
(red {line}) from Solutions 8, 11, and 12}
\label{RV}
\end{figure}
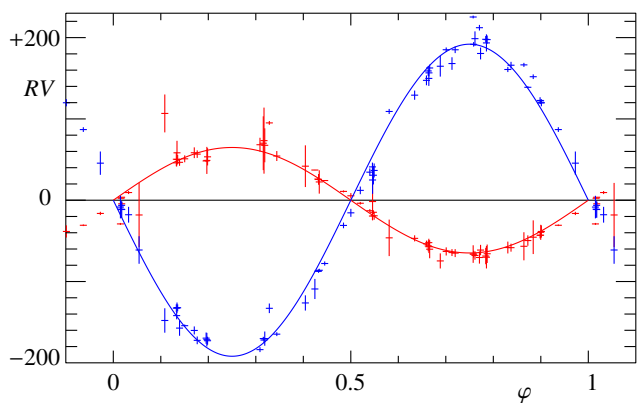

\begin{figure} 
\setlength{\unitlength}{1mm}
\begin{picture}(150,54)
\put(0,0){\includegraphics[width=\hsize]{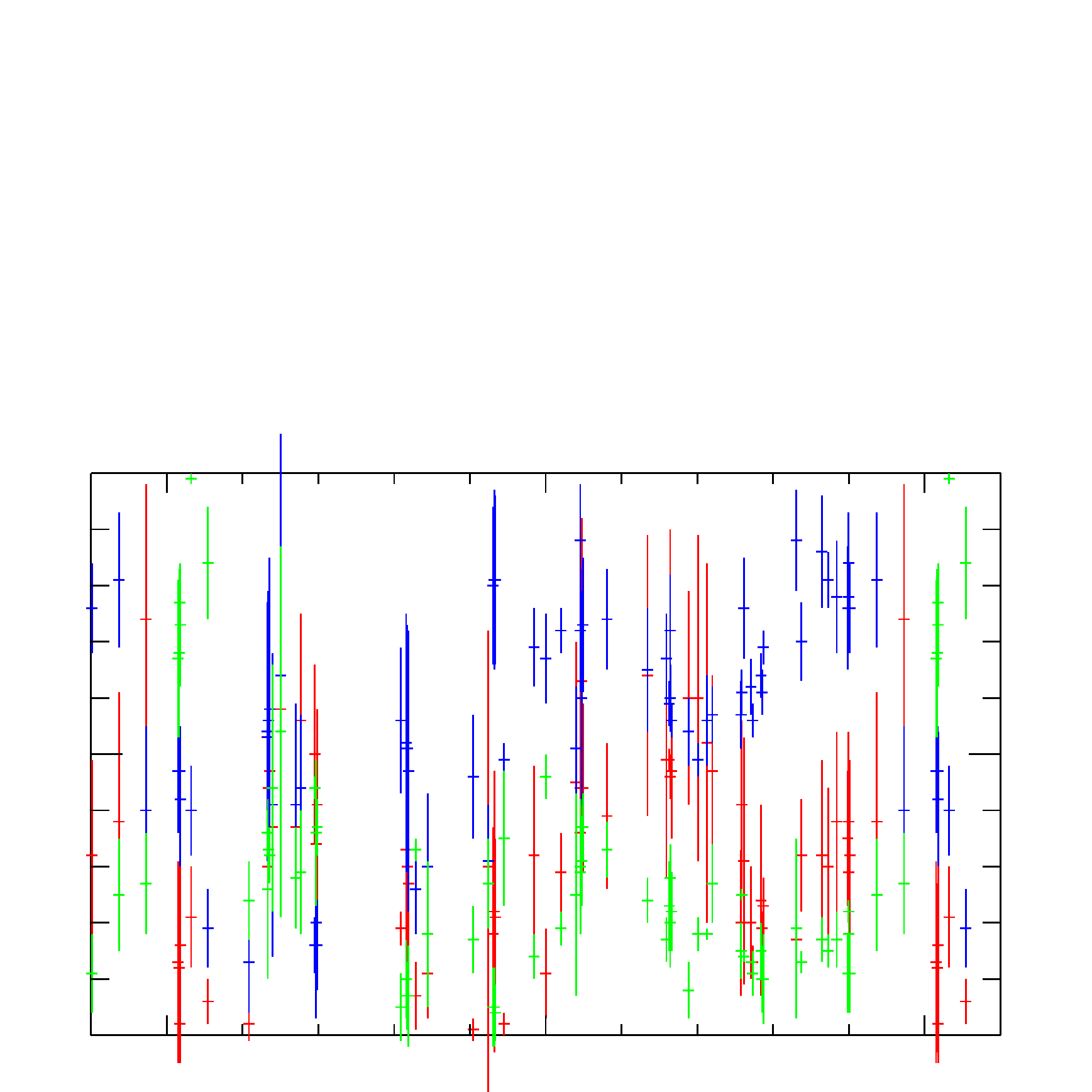}}
\put(3,27){\small 0.5}
\put(5,50){\small $1$}
\put(5,3.5){\small $0$}
\put(13,1){\small 0}
\put(43,1){\small 0.5}
\put(76,1){\small 1}
\put(70,0){\small $\varphi$}
\put(1,44){\small $e^s$}
\end{picture}
\caption{ {Phase dependence of exponential} of $s$-factors of donor (blue {marks}), gainer (red {marks}), and circumstellar component (green {marks}) from Solutions 8, 11, and 12}
\label{sf}
\end{figure}
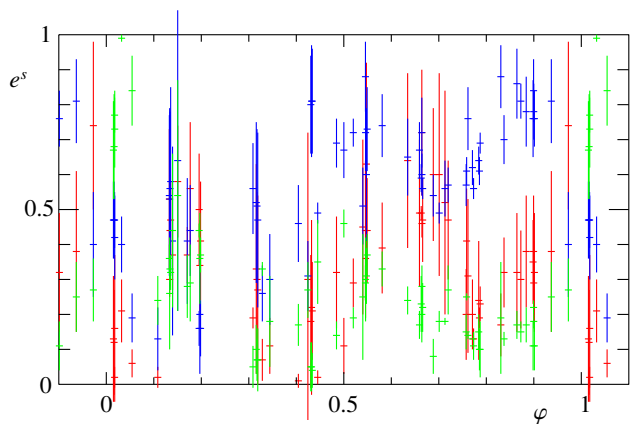

 For the 68 exposures from OAN, KAO, and AIO we have disentangled RVs and $s$-factors in all three above mentioned spectral regions of He\,I lines, which enables us to calculate the mean RVs and \lc{line-strength factors} from Solutions 8, 11, and 12.
The $s$-factors were normalised to zero for the strongest line in each solution.
Because the line strengths of some lines in some exposures are very small and, hence, their logarithmic $s$-factors are very negative, we computed the mean line strengths from the exponentials of $s$-factors, which are proportional to the depth in line flux.
In these exposures (those with extremely shallow lines) the RVs of the corresponding components are also poorly determined by the spectra, so we computed mean RVs with weights given by the exponentials of $s$-factors.
The resulting mean values and their errors from the He\,I lines 5875, 6678, and 7065\,{\AA} are given in Table~\ref{Tab1}.
The radial velocities are plotted in Fig.~\ref{RV}, where the RV curves correspond to the semi-amplitudes $K_d=191.9\,$km/s and $K_g=64.9\,$km/s computed as the mean values from the three solutions. 
The very {large} error-bars of RVs and some outliers correspond to exposures in which the line component has a small strength and hence the determination of the RV is unreliable.
The line strengths (on an exponential scale that is linear in flux) are shown in Fig.~\ref{sf}.

 The disentangling of spectral region around the line He\,I 4471\,{\AA} and the doublet of Mg\,II 4481\,{\AA} in OAN spectra sampled in 2048 bins with a step of 0.8 km/s is given in Solution 13. The line profile of He in the gainer is very shallow and the Mg line is almost entirely missing.

\subsection{Disentangling of nitrogen lines}
Spectral region 4990--5031\,\AA{} contains helium line He\,I 5015.7\,\AA{} and nitrogen lines N\,II 4994.4, 5001.5, 5005.1, and 5010.5 \AA{}.
Its disentangling shows the helium line of the donor, gainer, and the circumstellar component, while the nitrogen lines originate mostly from the donor and they are {poorly} distinguishable in the noise of the gainer and circumstellar component.
The disentangled orbital parameters are given in Solution 14.

 The spectral region 5656--5690\,\AA{} is dominated by nitrogen lines N\,II 5666.6, 5676.0, 5679.6, and 5686.2\,\AA,{} which are clearly disentangled in the spectrum of donor.
They are also distinguishable in the circumstellar component but no lines are disentangled in the spectrum of the donor.
The amplitude of its RVs converges to unrealistically small values. 
We thus disentangled this region in Solution 15 as a superposition of spectra of the donor and the circumstellar component only.

\begin{table*}
\caption{Disentangled orbital parameters of UU~Cas}\label{Tab0} 
\begin{tabular}{lcccccccccc}
Solution&            1&               2&               3&               4&               5&               6&               7\\ 
data    &          OAN&             OAN&             OAN&             KAO&            RNAO&             AIO&         GAO+APO\\
region  &   5860--5884&      5860--5884&      5860--5884&      5860--5884&      5860--5884&      5860--5884&      5860--5884\\
$T_d$ &$2.314\pm0.010$& $2.315\pm0.007$& $2.335\pm0.004$& $2.393\pm0.004$& $2.314\pm0.010$& $2.360\pm0.005$& $2.429\pm0.015$\\
$K_d$   &$191.5\pm1.4$& $191.6\pm1.0$  &   $186.6\pm0.8$&   $192.1\pm0.5$&   $200.6\pm2.0$&   $170.9\pm0.8$&   $192.5\pm1.8$\\
$Q$     &             &                &   $1.85\pm0.04$&   $2.25\pm0.03$&   $2.63\pm0.24$&   $2.26\pm0.04$&   $2.98\pm0.21$\\
$K_g$   &         <0.1&    $-8.4\pm2.2$&   $100.7\pm1.9$&    $85.4\pm1.0$&    $76.2\pm7.0$&    $75.5\pm1.2$&    $64.5\pm4.4$\\
$\sigma$&       0.0198&          0.0197&          0.0149&          0.0246&          0.0106&          0.0284&          0.0186\\ \hline
 Solution&               8&               9&              10&              11&              12&              13&              14\\ 
 data    &             all&             OAN&             AIO&             all&     OAN,KAO,AIO&             OAN&             OAN\\
 region  &      5860--5884&      6662--6691&      6662--6691&      6662--6691&      7054--7078&      4462--4486&      4990--5031\\
 $T_d$   & $2.391\pm0.006$& $2.298\pm0.007$& $2.297\pm0.003$& $2.325\pm0.008$& $2.319\pm0.009$& $2.063\pm0.008$& $2.300\pm0.010$\\
 $K_d$   &   $187.6\pm0.9$&   $190.2\pm1.1$&   $187.0\pm0.4$&   $194.1\pm1.1$&   $195.2\pm1.7$&   $198.0\pm2.0$&   $197.0\pm1.6$\\
 $Q$     &   $2.81\pm0.10$&   $2.11\pm0.04$&   $2.08\pm0.02$&   $3.39\pm0.18$&   $3.56\pm0.19$&   $2.29\pm0.05$&   $3.32\pm0.17$\\
 $K_g$   &    $66.9\pm2.4$&    $90.1\pm1.6$&    $90.1\pm0.9$&    $57.3\pm3.0$&    $54.9\pm2.9$&    $86.6\pm1.9$&    $59.4\pm3.0$\\
$\sigma$ &          0.0249&          0.0196&          0.0213&          0.0226&          0.0253&          0.0276&          0.0162\\
\hline 
Solution&              15&              16&              17&              18&              19&              20& mean\\ 
 data   &             OAN&             OAN&             OAN&             OAN&             AIO&             KAO& all\\
 region &      5656--5690&      4312--4365&      4840--4889&      6540--6607&      6540--6607&      6540--6580& \\
 $T_d$  & $2.266\pm0.014$& $2.063\pm0.007$& $2.366\pm0.009$& $2.503\pm0.036$& $2.457\pm0.016$& $2.280\pm0.007$& $2.319\pm0.087$\\
 $K_d$  &   $196.8\pm2.0$&   $193.4\pm1.1$&   $189.6\pm1.7$&   $129.7\pm6.2$&   $155.8\pm2.9$&   $169.7\pm2.2$& $191.0\pm7.4$\\
 $Q$    &                &   $2.63\pm0.04$&   $2.96\pm0.14$&   $2.32\pm0.14$&   $2.71\pm0.09$&   $3.34\pm0.08$& $2.54\pm0.68$\\
 $K_g$  &                &    $73.5\pm1.2$&    $64.1\pm3.0$&    $55.9\pm4.4$&    $57.5\pm1.9$&    $50.8\pm1.3$& $75.6\pm14.4$\\
$\sigma$&          0.0152&          0.0307&          0.0193&          0.0223&          0.0224&          0.0340& \\
\end{tabular}
\tablefoot{
$T_{d}$ means HJD--2443130 of {the eclipse of donor; $K_d$, $K_g$ semiamplitudes in km\,s$^{-1}$ of the RV-curves of the donor and gainer, respectively (note that $K_g$ is suppressed by the circumstellar component in the two-component Solutions 1 and 2); $Q$ mass ratio}; $\sigma$ residual noise.
}

\end{table*}


\subsection{Disentangling of interstellar Na doublet and DIBs}
In order to check and, possibly, to improve the calibration of the wavelength-scale in the individual exposures, we disentangled the spectral region 5885--5899\,{\AA} sampled in 2048 bins (i.e. with a step 0.35\,km/s), which contains the Na doublet D1 (5895.923\,\AA) and D2 (5889.953\,\AA).
Both these interstellar absorption lines have doubled structures with bottoms at about 18\% of the continuum level.
Their fit by a superposition of Gaussian profiles in OAN spectra reveals that the weaker components are blue-shifted for $-43.78\pm0.67$\,km/s and the stronger components for $-1.22\pm0.61$\,km/s.
The mean half-width of these lines corresponds to approximately 21\,km/s.
The original mean scatter of RVs of these interstellar lines was measured from the independently reduced OAN spectra (of about 3\,km/s) was reduced to 0.01\,km/s after applying the correction to the input spectra.

 The scatter of RVs of Na D lines in AIO spectra was improved from 0.5\,km/s to 0.01\,km/s.
It is interesting to note that the low-velocity components of the Na doublet are contaminated by an emission from the street lamps in Prague backscattered on thin clouds in some exposures.
The airglow emission lines of O\,I (in particular at 5577 and 6300 {\AA}) also appear in these exposures.
Because the input spectra are transformed to the heliocentric wavelength scale, these emissions share the annual motion of the telluric lines.
Their strengths depend on the weather conditions and must be \lc{found numerically in order to enable the disentangling of these lines.}

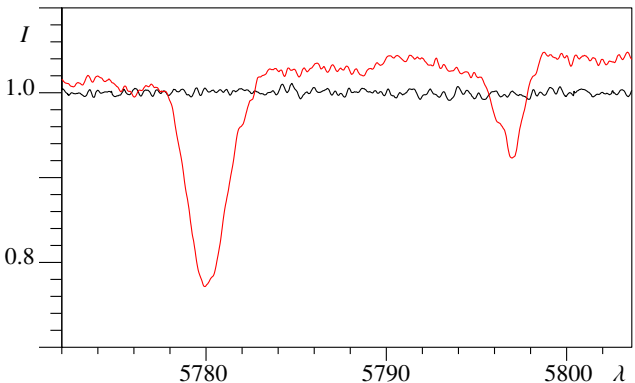
\begin{figure} 
\setlength{\unitlength}{1mm}
\begin{picture}(150,50)
\put(0,-3){\includegraphics[width=\hsize]{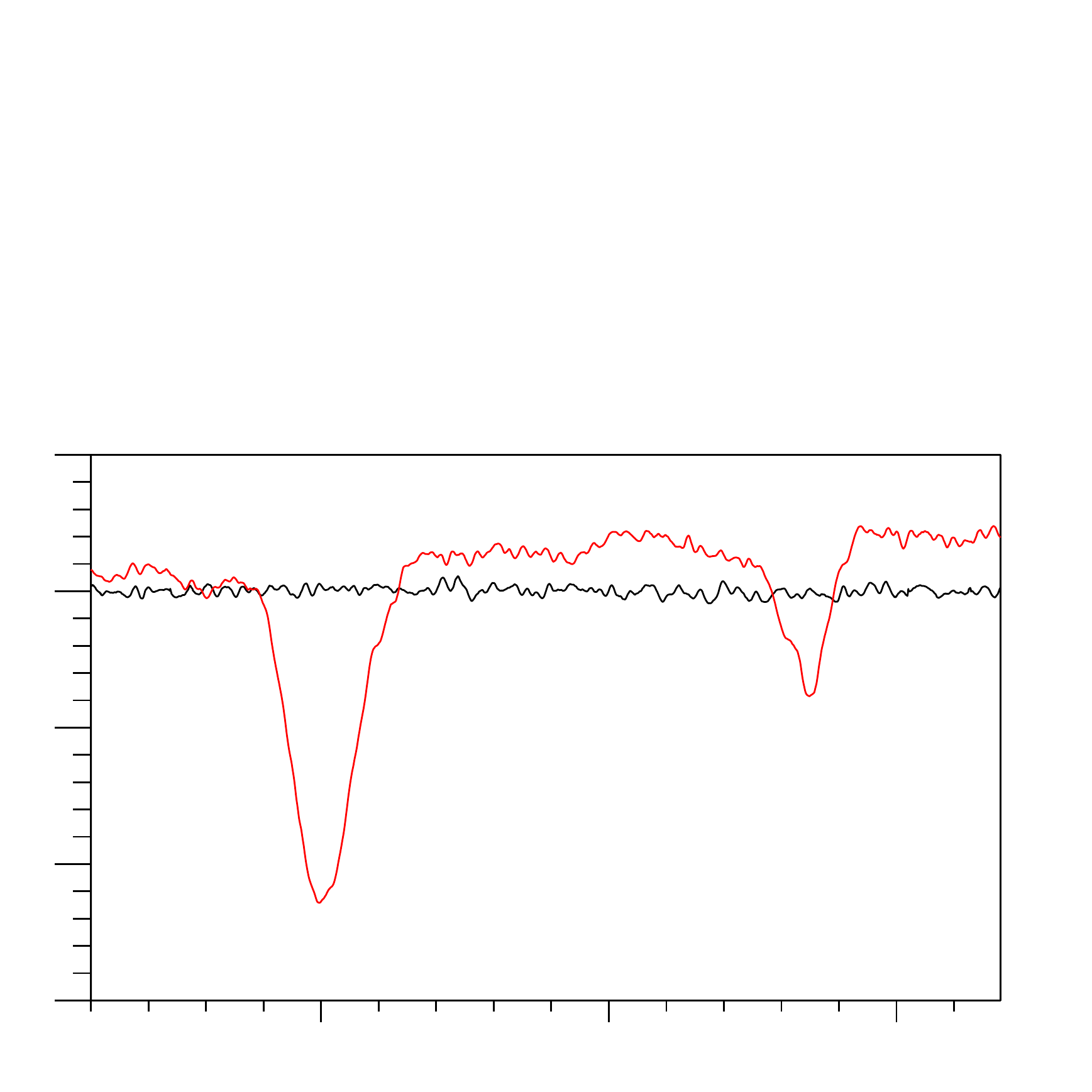}}
\put(2,45){\small $I$}
\put(0,38){\small 1.0}
\put(0,15.5){\small 0.8}
\put(23,0){\small 5780}
\put(47,0){\small 5790}
\put(71,0){\small 5800}
\put(80,0){\small $\lambda$}
\end{picture}
\caption{Disentangled featureless spectrum of the donor (black line) and DIBs (red line) in the region 5772--5803\,\AA{} in the OAN spectra }
\label{P5772}
\end{figure}

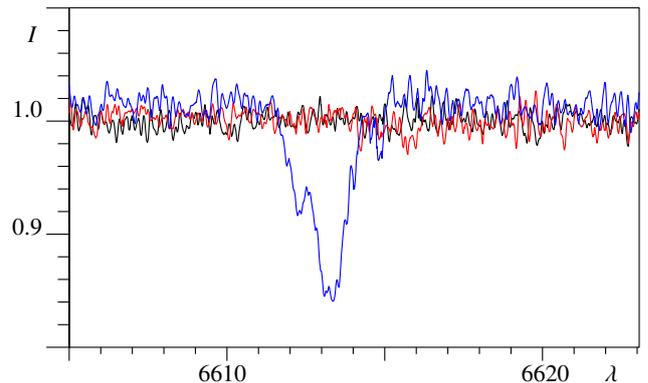
\begin{figure} 
\setlength{\unitlength}{1mm}
\begin{picture}(150,50)
\put(2,45){\small $I$}
\put(0,-3){\includegraphics[width=\hsize]{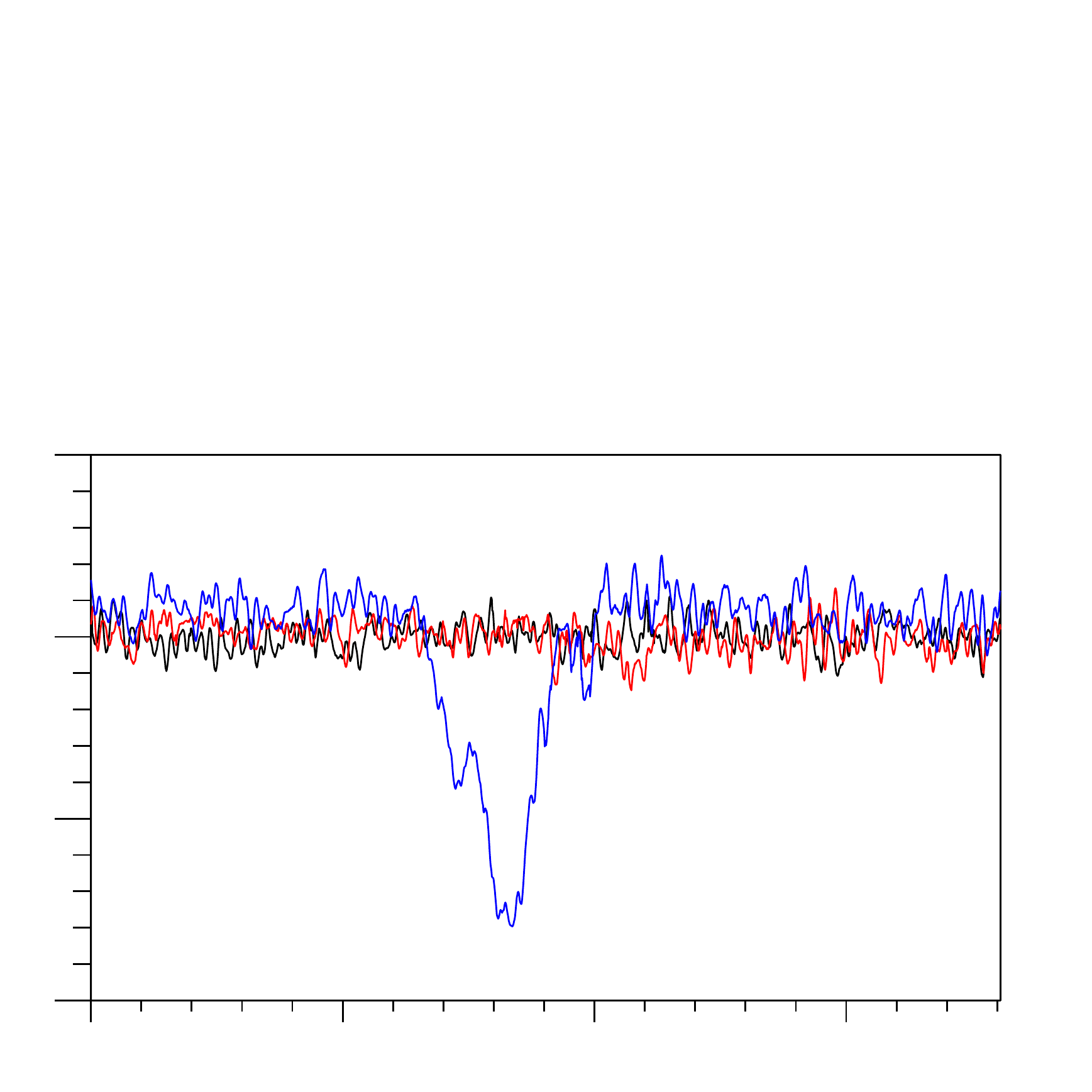}}
\put(0,35){\small 1.0}
\put(0,19.4){\small 0.9}
\put(24.5,0){\small 6610}
\put(67,0){\small 6620}
\put(78,0){\small $\lambda$}
\end{picture}
\caption{Disentangled featureless spectra of the donor (black line), gainer (red line), and DIBs (blue line) in the region 6605--6623\,\AA{} in the AIO spectra }
\label{O6605}
\end{figure}

 \lc{We also} disentangled diffuse interstellar bands (DIBs) 5780.5, 5797.1, and 6613.6 \AA{} \citep[cf.][Figs~\ref{P5772}, \ref{O6605}]{2013ApJ...773...42O}. 
Unlike the circumstellar lines, the strength of DIBs does not change with the orbital phase.

\subsection{Disentangling of Balmer lines}
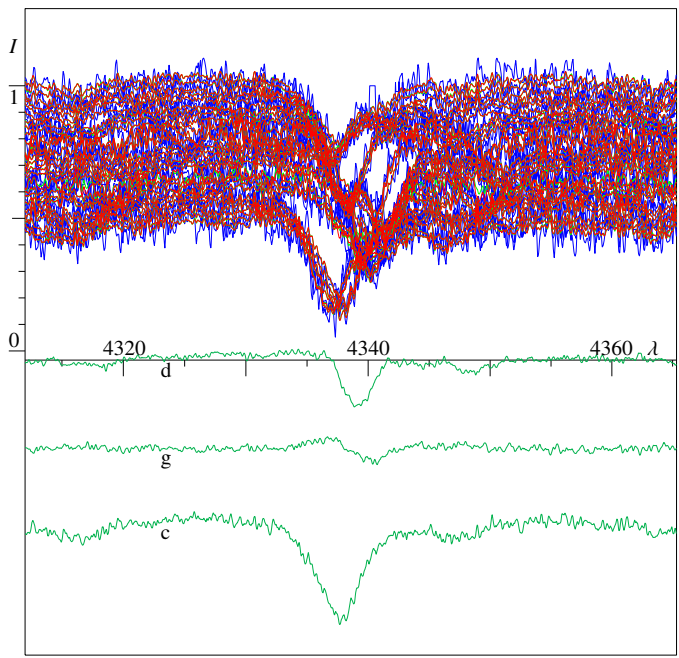
\begin{figure} 
\setlength{\unitlength}{1mm}
\begin{picture}(150,90)
\put(0,0){\includegraphics[width=\hsize]{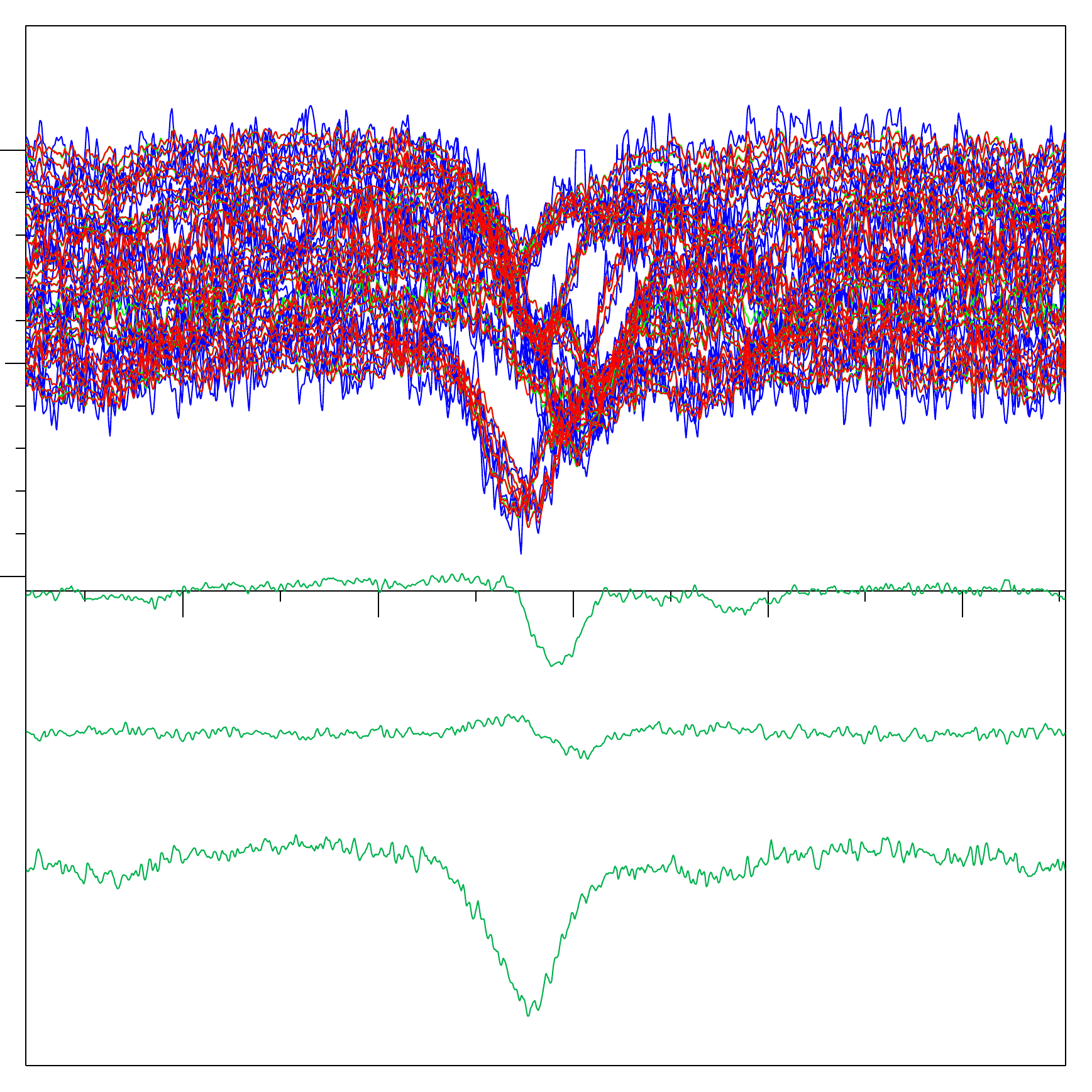}}
\put(0,82){\tiny $I$}
\put(0,75.2){\tiny 1}
\put(0,43){\tiny 0}
\put(12.4,42){\tiny 4320}
\put(44.5,42){\tiny 4340}
\put(76.5,42){\tiny 4360}
\put(84,42){\tiny $\lambda$}
\put(20,39){\tiny d}
\put(20,27.5){\tiny g}
\put(20,17.5){\tiny c}
\end{picture}
\caption{Disentangled profiles of H$\gamma$ line of UU~Cas from OAN spectra (Solution 16)}
\label{P4312}
\end{figure}

\begin{figure} 
\setlength{\unitlength}{1mm}
\begin{picture}(150,90)
\put(0,0){\includegraphics[width=\hsize]{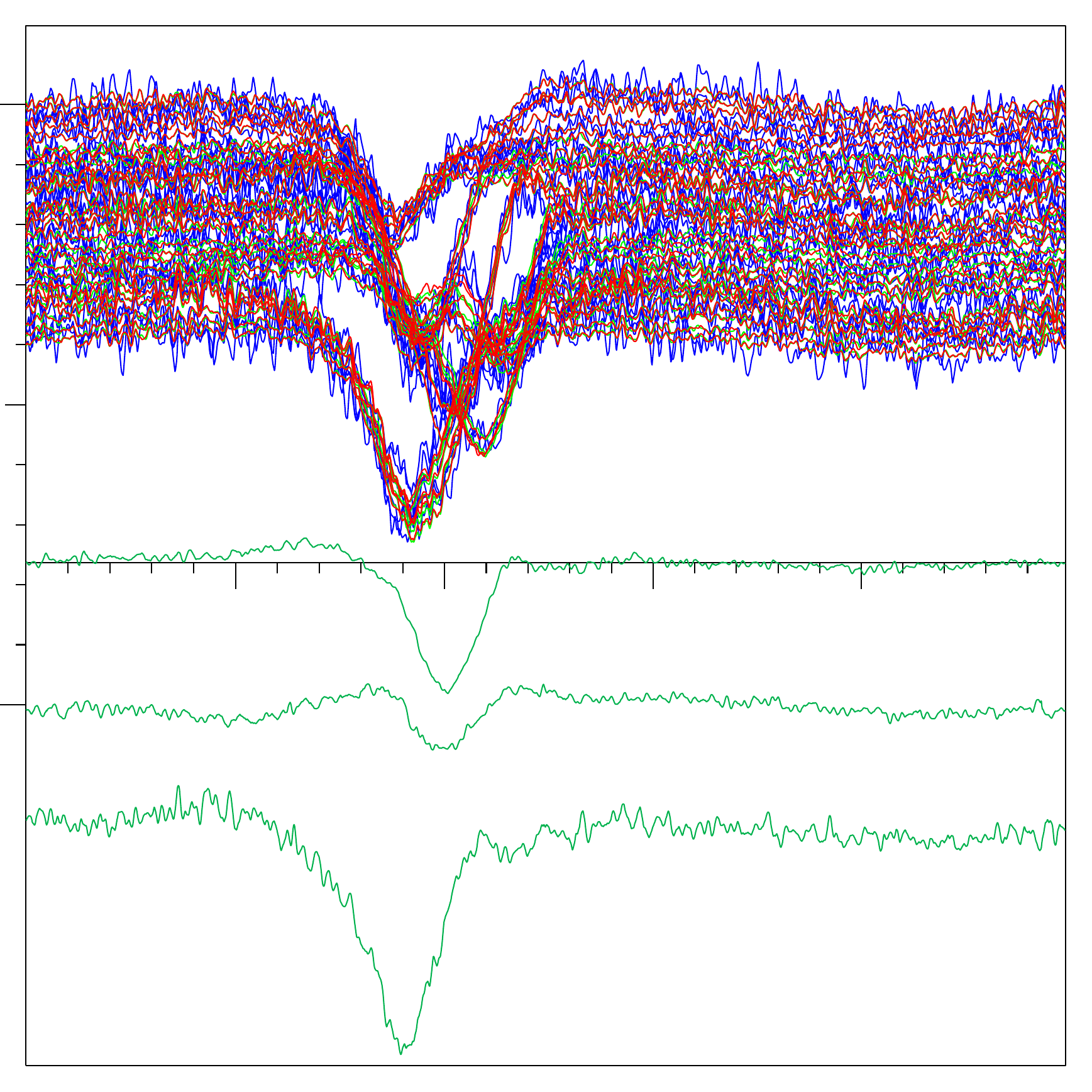}}
\put(0,83){\tiny $I$}
\put(0,79){\tiny 1}
\put(0,32.5){\tiny 0}
\put(16.5,44){\tiny 4850}
\put(68.5,44){\tiny 4880}
\put(83,44){\tiny $\lambda$}
\put(25,45.8){\tiny d}
\put(25,33.4){\tiny g}
\put(25,20.5){\tiny c}
\end{picture}
\caption{Disentangled profiles of H$\beta$ line of UU~Cas from OAN spectra (Solution 17)}
\label{P4840}
\end{figure}
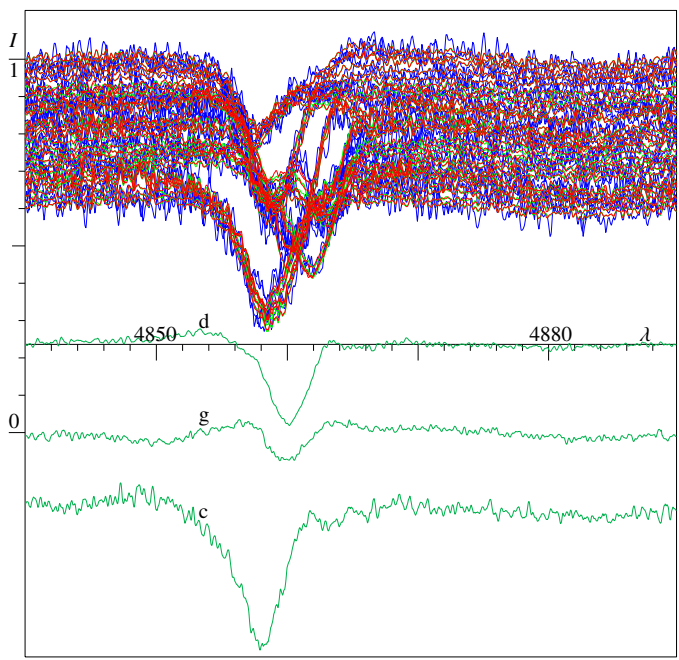

The  H$\gamma$ line can be disentangled into the three components, shown in Solution 16 and Fig.~\ref{P4312}.
The line profile of the gainer has faint emission wings. 
This feature is much more pronounced in the disentangling of H$\beta$ line (cf. Solution 17 and Fig.~\ref{P4840}).
{It was pointed out by H. Hensberge in his referee report that the region with H$\beta$ line also contains 
a broad DIB \citep[at 4881.83\,{\AA}, FWHM 19.67\,{\AA}, cf.][]{1994A&AS..106...39J}, which can be recognised in 
the disentangled spectra of the circumstellar component and, partly,  also that of the gainer, both of which have no or small radial 
velocities.
In principle, the profile of the DIB could also be disentangled as an independent component because its strength is invariable, unlike the other components.
Such a free disentangling of the fourth interstellar component effectively flattened the continua of the gainer and the circumstellar component on the long-wavelength side of H$\beta$ and moved the wide shallow depression into the constant interstellar component; however, at the same time, it redistributed a part of the H$\beta$ emission into this component.
This effect is understandable because the profiles of the three components do not change in strength only, but also a bit in shape and, hence, the new degrees of freedom in the fourth component are used to fit the observed
profiles better.
This can be avoided using the template-constrained disentangling, that is, if we predefine the shape of the interstellar
component by a synthetic profile or by the result of the free disentangling with the distortions \lc{in the profile of} H$\beta$ removed.
In practice, this procedure applied to the current spectra of UU~Cas did not bring any reliable improvement of the solution, apart from the flattening of the continua.
Nevertheless, it may prove to be useful if in the future when more precise observations will enable better determinations of the profile of the DIB.
}

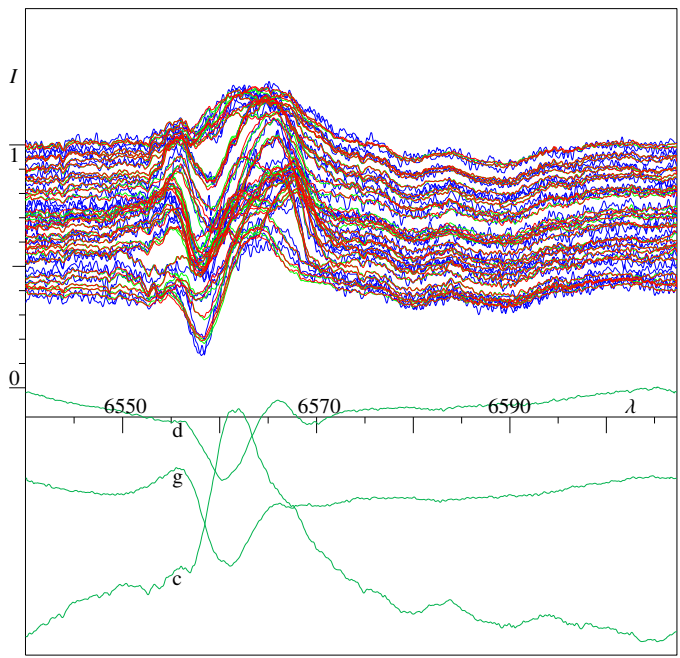
\begin{figure} 
\setlength{\unitlength}{1mm}
\begin{picture}(150,90)
\put(0,0){\includegraphics[width=\hsize]{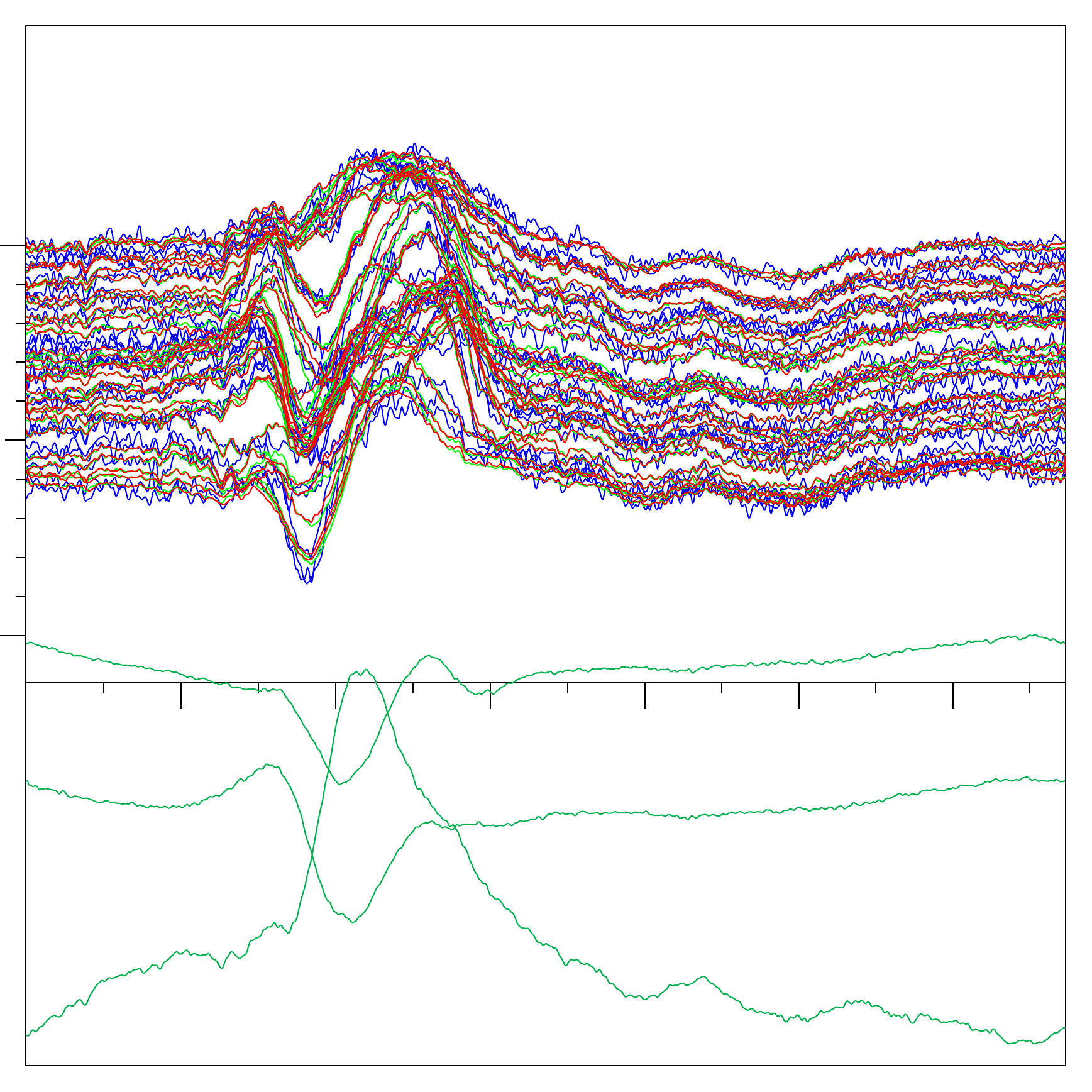}}
\put(0,78){\tiny $I$}
\put(0,67.5){\tiny 1}
\put(0,38){\tiny 0}
\put(12.5,34.3){\tiny 6550}
\put(38,34.3){\tiny 6570}
\put(63,34.3){\tiny 6590}
\put(81,34.3){\tiny $\lambda$}
\put(21.5,31){\tiny d}
\put(21.5,24.9){\tiny g}
\put(21.5,11.5){\tiny c}
\end{picture}
\caption{Disentangled profiles of H$\alpha$ line of UU~Cas from OAN spectra (Solution 18)}
\label{P6540}
\end{figure}

\begin{figure} 
\setlength{\unitlength}{1mm}
\begin{picture}(150,90)
\put(0,0){\includegraphics[width=\hsize]{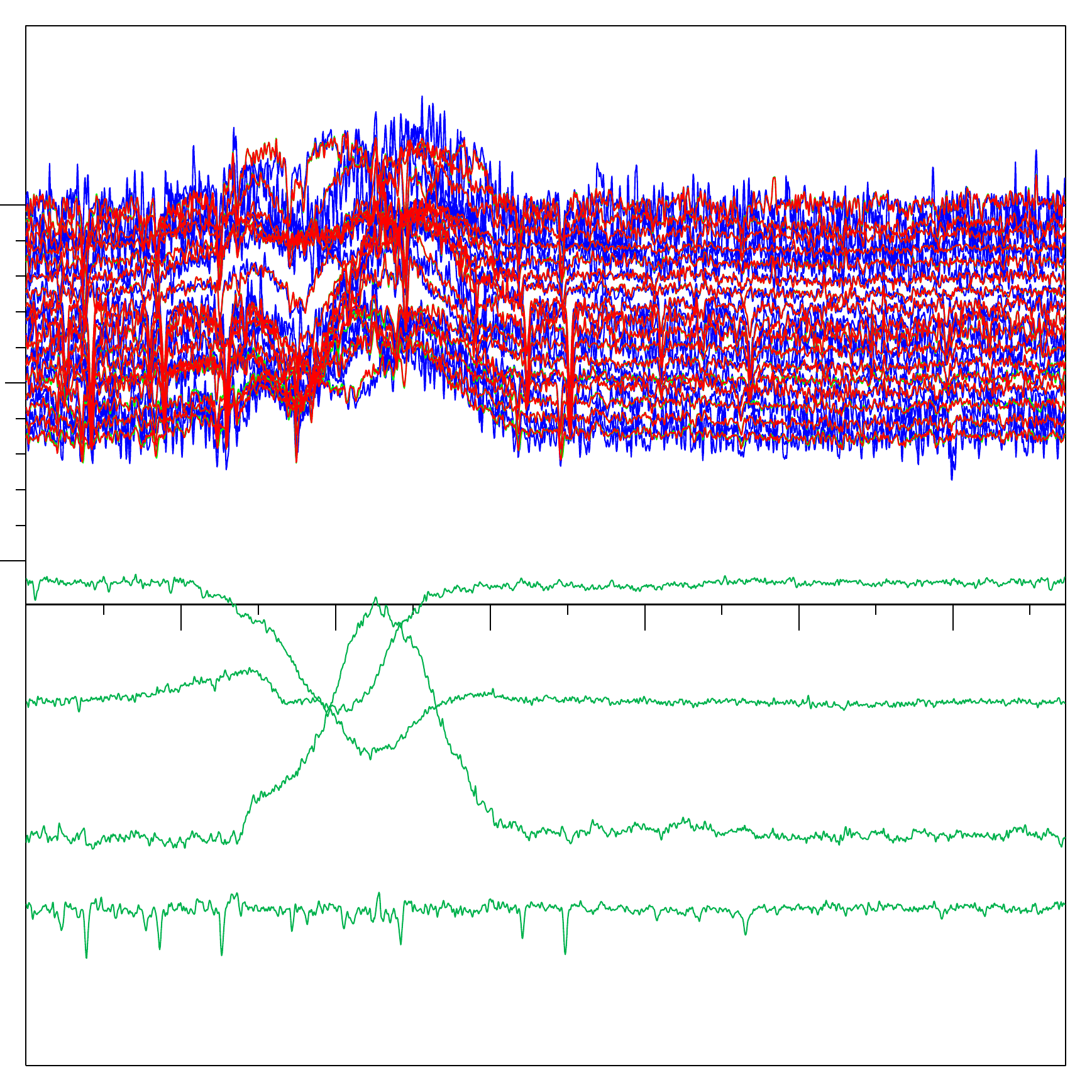}}
\put(0,80){\tiny $I$}
\put(0,73.7){\tiny 1}
\put(0,44.2){\tiny 0}
\put(12.5,40.8){\tiny 6550}
\put(38,40.8){\tiny 6570}
\put(63,40.8){\tiny 6590}
\put(81,40.8){\tiny $\lambda$}
\put(20,36.6){\tiny d}
\put(20,32.3){\tiny g}
\put(20,19){\tiny c}
\put(20,12.5){\tiny t}
\end{picture}
\caption{Disentangled profiles of H$\alpha$ line of UU~Cas from AIO spectra (Solution 19).
 The fourth bottom line is the disentangled telluric spectrum (water-vapour lines)}
\label{O6540}
\end{figure}
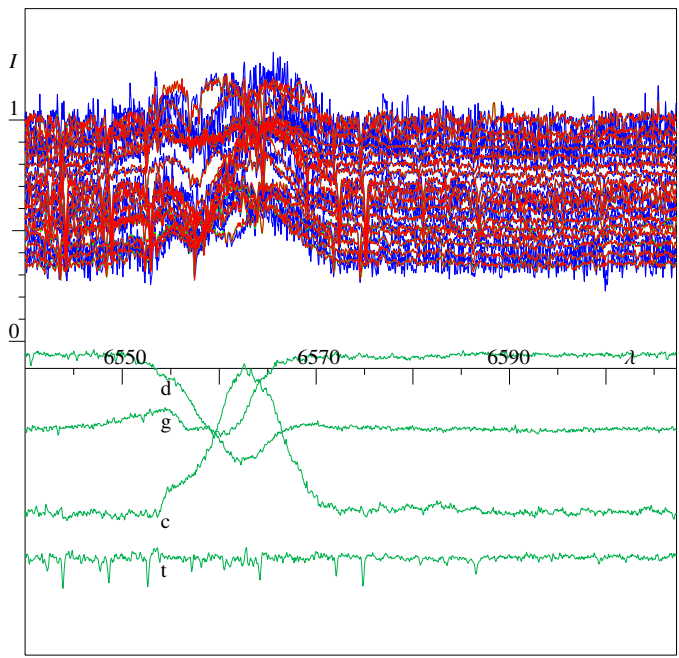

\begin{figure} 
\setlength{\unitlength}{1mm}
\begin{picture}(150,90)
\put(0,0){\includegraphics[width=\hsize]{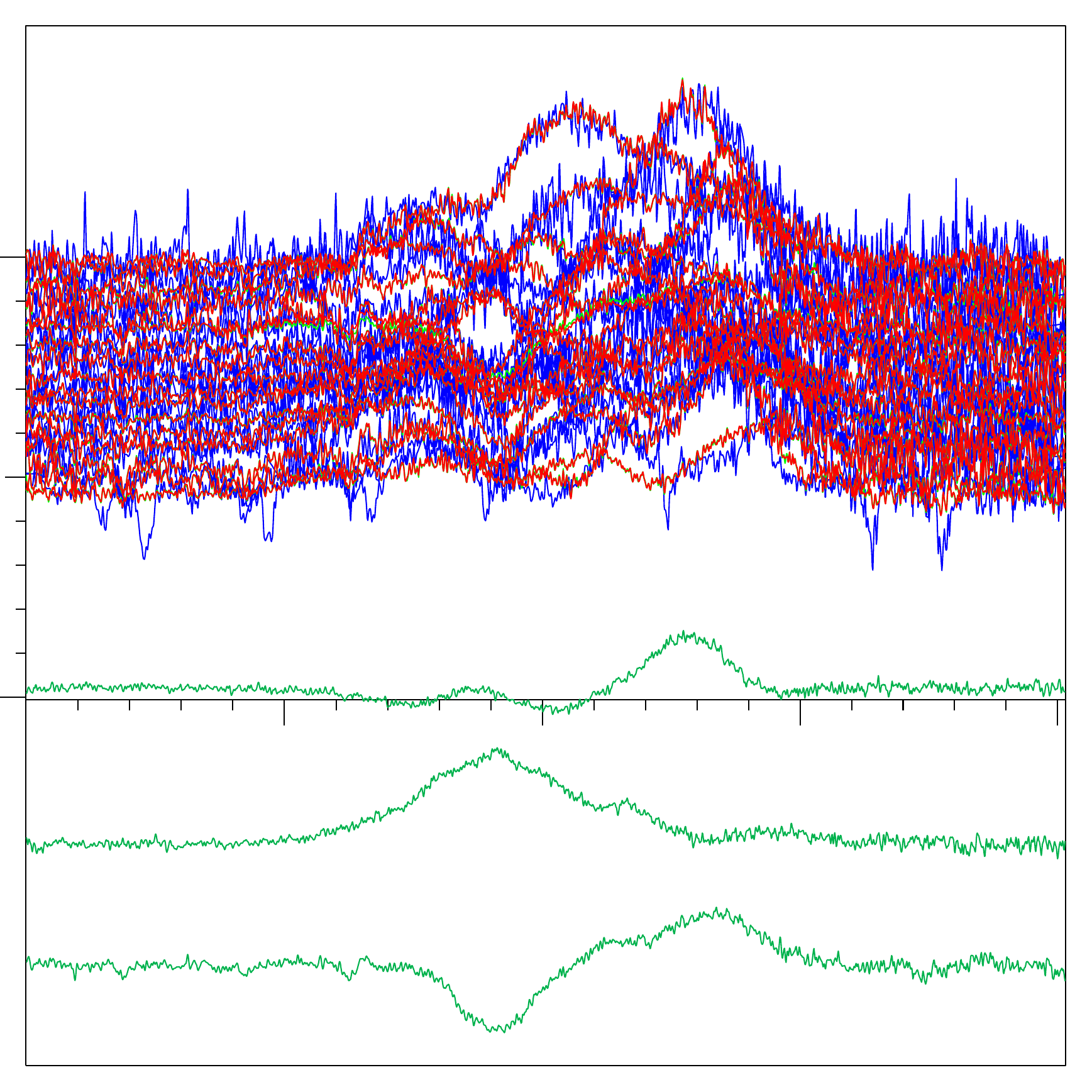}}
\put(0,80){\tiny $I$}
\put(0,69.2){\tiny 1}
\put(0,33.){\tiny 0}
\put(20.7,33.4){\tiny 6550}
\put(63.3,33.4){\tiny 6570}
\put(81,33.4){\tiny $\lambda$}
\put(12,33.9){\tiny d}
\put(12,22){\tiny g}
\put(12,11.5){\tiny c}
\end{picture}
\caption{Disentangled profiles of H$\alpha$ line of UU~Cas from KAO spectra of 2017 (Solution 20).}
\label{K6540}
\end{figure}
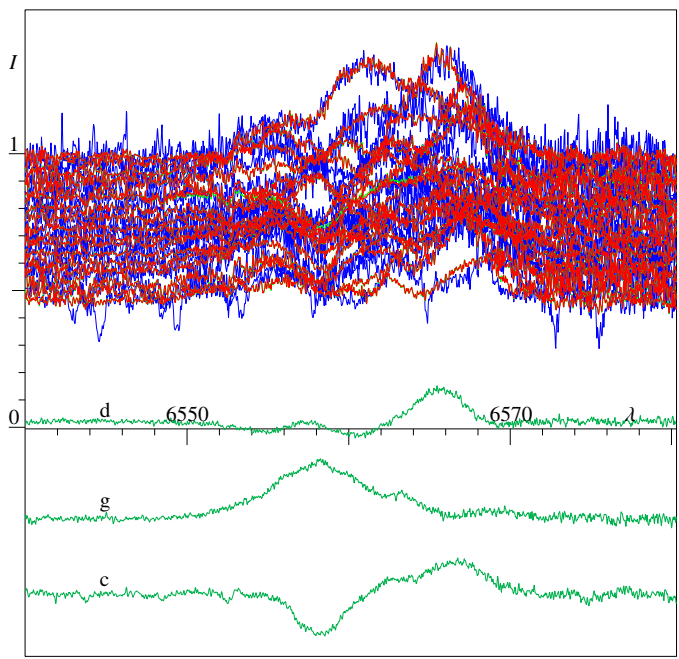

 Finally, the H$\alpha$ profile can also be disentangled to the three components.
In the OAN spectra (cf. Solution 18 and Fig.~\ref{P6540}) and AIO spectra (Solution 19 and Fig.~\ref{O6540}), we find the circumstellar component in emission instead of in the absorption as it is in the higher Balmer lines.
While the line profiles of the donor and gainer in both these data-sets are mostly in absorption with a moderate emission features only, the KAO spectra yield a significantly different picture which, moreover, changes between the observational seasons 2017 and 2018.
Both stellar components appear to be in emission in 2017, while the circumstellar component has an absorption core with a red-shifted emission peak (cf. Solution 20 and Fig.~\ref{K6540}).
In 2018, all three components have enhanced red emission wings separated only by  shallow absorptions from the moderate blue wings.
Nine RNAO spectra obtained in 2009 and 2010 do not  sufficiently cover the second half of the orbital period, so their disentangling is unreliable.
Nevertheless, these spectra indicate yet another structure of the components's contributions to the overall H$\alpha$ profile, with the red-shifted emission of the gainer and double-peaked emission of the circumstellar component.
These changes in the line profiles, together with variations of the disentangled values of $K$-velocities, indicate an irregular behaviour on the part of the circumstellar matter in the system, to which the H$\alpha$ line is particularly sensitive.

 To check a possible correlation between the changes in the structure of the circumstellar matter and the long photometric period 269 days found by \citet{2020A&A...642A.211M}, 
we performed a Fourier analysis of equivalent widths of the H$\alpha$ line.
The periodogram showed a number of not much pronounced local maxima, the strongest one at the period 4.25 days, namely, half of the orbital period, the next strongest one at 346.6 days, but none at 269 days.
This may, however, be influenced by the \lc{insufficient coverage of all phases of the orbital period in all phases of the long period by the currently available spectroscopic observations}.

\subsection{Orbital parameters}
The orbital parameters disentangled from the H$\alpha$ \lc{line} and, partly, also the H$\beta$ and H$\gamma$ lines, have a larger scatter than those found from the He\,I lines.
This confirms the above-mentioned assumption that the Balmer lines are not suitable for a determination of the orbital parameters.
We thus calculated mean values of the parameters from the Solutions 3--14, namely, those based on the He lines only (cf. the last solution in Table~\ref{Tab0}).
The errors of these mean parameters are larger than the mean errors of the individual solutions, which is partly due to the mentioned underestimation of the Bayesian errors which neglect the adaptability of the solutions enabled by the \lc{line-strength factors}.
The errors of the spectroscopic orbital parameters may, however, also be significantly increased due to the \lc{effect} of aliasing the RVs of component stars by the circumstellar matter, which differs according to the various seasons, phases, and lines.

 As noticed by H. Hensberge, there seems to be a systematic difference between the smaller value of the mass ratio $Q$, which is disentangled from individual sets of spectra treated separately, and the higher value of $Q$ obtained when more sets are disentangled together.
Indeed, this is true not only for the examples of Solutions 3, 4, and 6 compared to Solution 8 or Solutions 9 and 10 -- as compared to 11 -- but also for others, which we left out of Table~\ref{Tab0} for brevity.
For instance, the OAN and AIO spectra give $Q=2.70$ and 2.71, respectively, for the region 7054--7078\,\AA{} compared to 3.56 in Solution 12.
If we merge together the OAN and AIO spectra used for Solutions 9 and 10, via their disentangling we obtain $Q=2.70$, which is also significantly higher than the mutually consistent values found from these sets separately.
This is because the line profiles of the circumstellar component disentangled from these sets separately have different shapes and the $\gamma$-velocities of both components also differ, which indicates a change in the \lc{mass-overflow rate} between the two periods of observation.
Thus, \lc{the common solution does not fit} both sets comparably well (its residual noise is $\sigma=0.0229$).
If the smoothing \lc{of the input spectra for} different states of mass transfer leads to an increase in $Q$, it can also explain its higher value in Solution 5 of RNAO spectra, which have been collected over the longest observational period, and the Solution 7, for which templates of the gainer and the circumstellar component were taken from Solution 3.
\lc{If these questionable solutions are omited, the averaging over Solutions 3, 4, 6, 9, 10, 13, and 14 results} in $Q=2.31\pm0.44$, if the above-mentioned solutions of OAN and AIO spectra at 7054--7078\,\AA{} are \lc{also included, a mean value of $Q=2.40\pm0.42$ is} obtained.
It is, however, not clear why the change in the \lc{mass-transfer rate} influences the observed value of $Q$ so much and which value is more realistic.
There are also great differences between different solutions for OAN spectra alone (cf. Solutions 3 and 14).
We thus keep the most conservative estimate of the mean solution given in Table~\ref{Tab0} until either new data or a more precise modelling of the observed line profiles provides a more reliable value of the mass ratio.

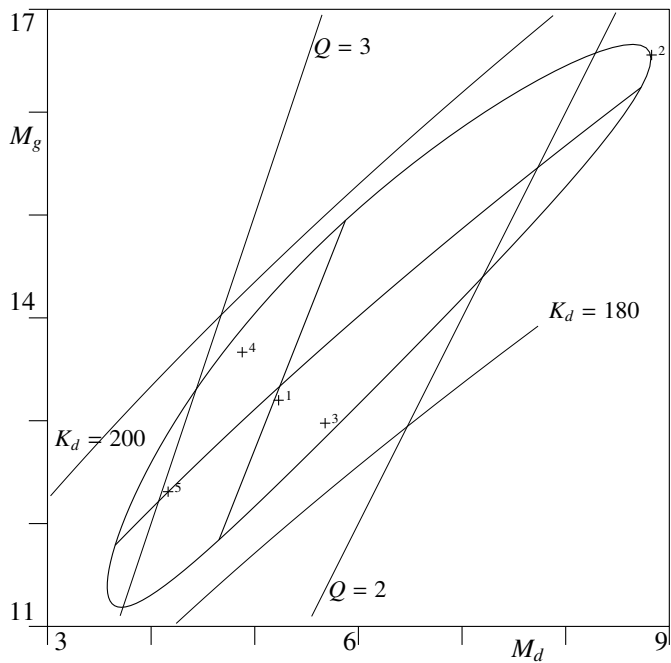
\begin{figure} 
\setlength{\unitlength}{1mm}
\begin{picture}(150,90)
\put(1,0) 
{\includegraphics[width=\hsize]{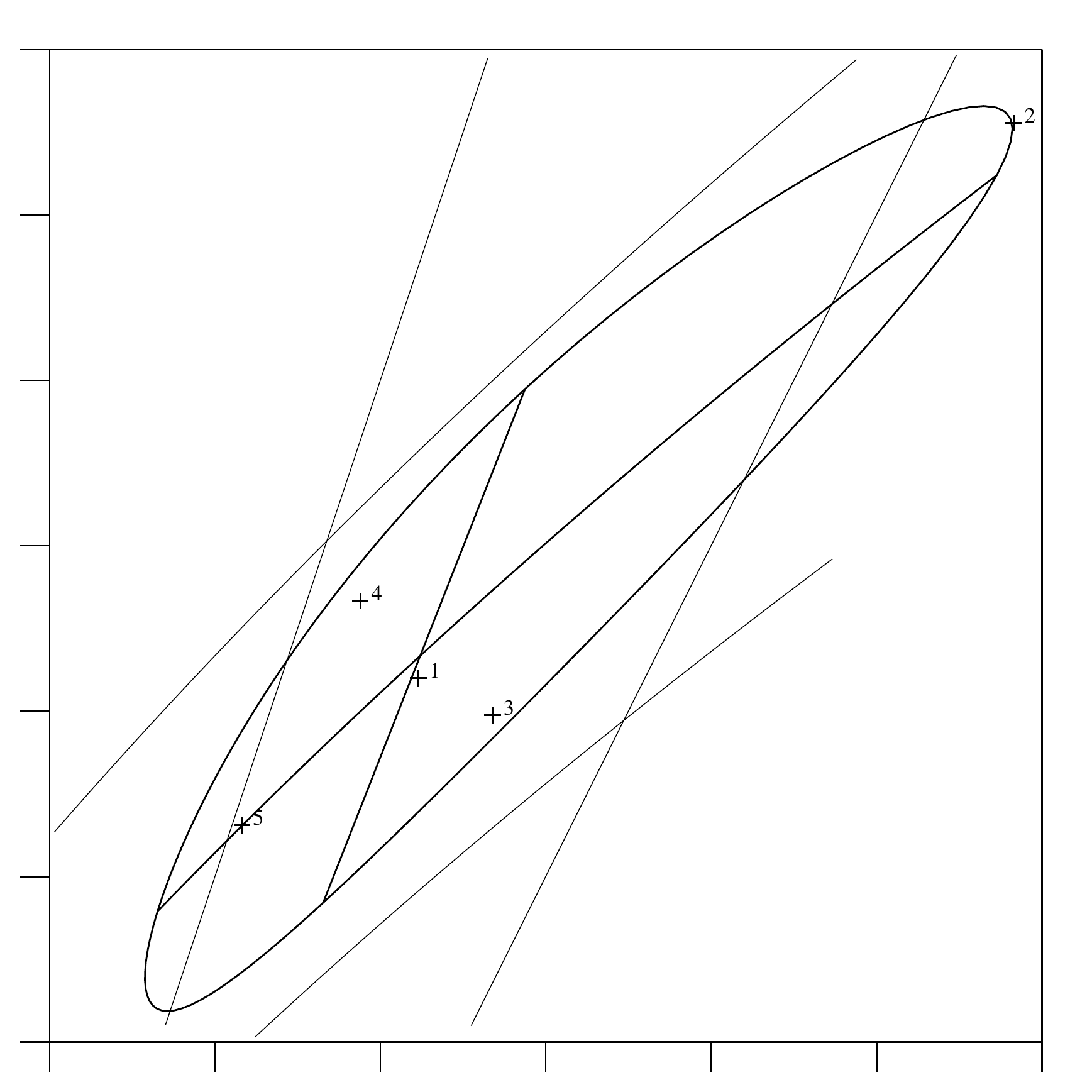}}
\put(66,0){$M_d$}
\put(6,1){3}
\put(44,1){6}
\put(85,1){9}
\put(0,68){$M_g$}
\put(0,5){11}
\put(0,46){14}
\put(0,83){17}
\put(71,45){\small $K_d=180$}
\put(6,28){\small $K_d=200$}
\put(40,80){\small $Q=3$}
\put(42,8){\small $Q=2$}
\end{picture}
\caption{Region of 1-$\sigma$ probability of component masses in units of $M_\odot$ corresponding to the mean solution ($M_d=5.24 M_\odot$, $M_g=13.33 M_\odot$) of the disentangled orbital parameters. The numbered marks denote the final masses of evolutionary models in Table \ref{Tabevol}}
\label{sols}
\end{figure}

 Accepting the inclination of the orbit $74.5^\circ$, the values of parameters $K_d$ and $Q$ result in masses of $M_d=5.24 M_\odot$ and $M_g=13.33 M_\odot$ for the component stars.
However, the wide error-ellipse of these parameters implies a large uncertainty in the masses, shown in Fig.~\ref{sols}.

\subsection{Parameters of the donor atmosphere}                              
The parameters of atmosphere of a binary component, namely, the effective temperature $T_{\rm eff}$, log~$g$, rotational broadening, and, possibly, the chemical composition can be estimated either by fitting of synthetic spectra to the disentangled spectrum of the component or by optimizing the residual noise in the disentangling constrained by templates constructed from the synthetic spectra.
In both cases, a fit of the wavelength shift due to the $\gamma$-velocity, fit of the rotational broadening, and scaling of the strength of lines due to the contribution of the other component stars to the overall continuum of the system must be performed.
In this study, we used the BSTAR2006 grid of synthetic spectra \citep[cf.][]{2007ApJS..169...83L} 
to construct the synthetic templates or to fit the disentangled spectra. 
The best fit was achieved with the models computed for the microturbulence velocity 10\,km/s and abundances enhanced by the CNO-cycle.
This is consistent with the evidence for the mass loss from the donor star yielded by the fading of this component at the phases shortly after the primary eclipse, which can be caused by the gaseous stream outflowing from the donor star at the vicinity of the point $L_1$.
 
\begin{figure*}
\includegraphics[width=\hsize]{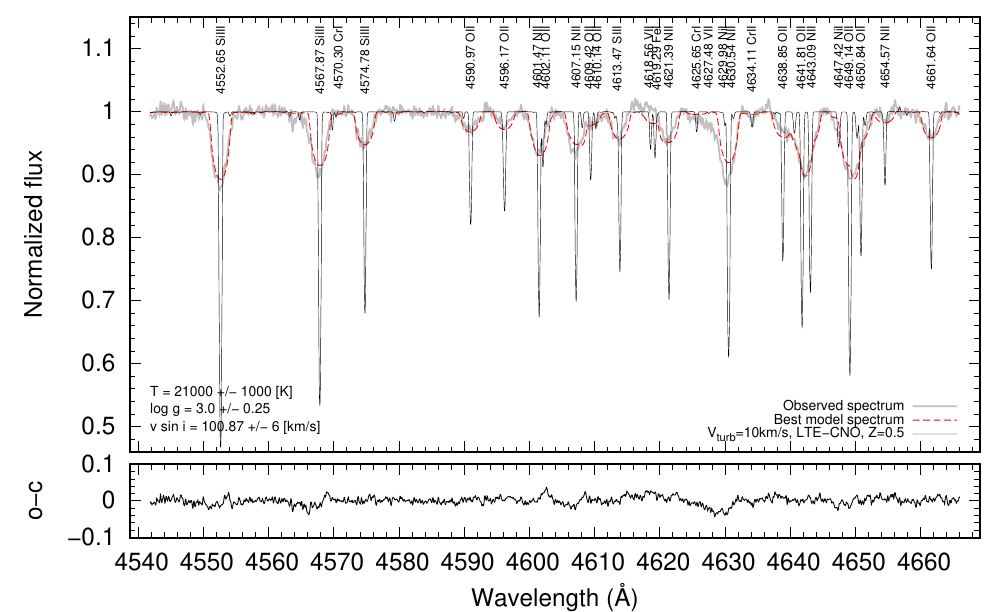} 
\caption{Fit of the disentangled donor spectrum in the region 4540--4665\,\AA{} in OAN spectra of UU~Cas}           
\label{Fit4540}                                                                  
\end{figure*}

 The parameters of the atmosphere are mainly constrained by ratios of equivalent widths of different spectral lines.
While a narrow spectral region sampled in a high resolution around a one or a few sharp spectral lines is preferable for the determination of the orbital parameters, a wide region with many lines is thus better  when establishing the atmosphere parameters.
The orbital parameters are fixed in these calculations to the values found in the narrow regions.

\begin{figure} 
\setlength{\unitlength}{1mm}
\begin{picture}(150,150)
\put(0,0){\includegraphics[width=\hsize]{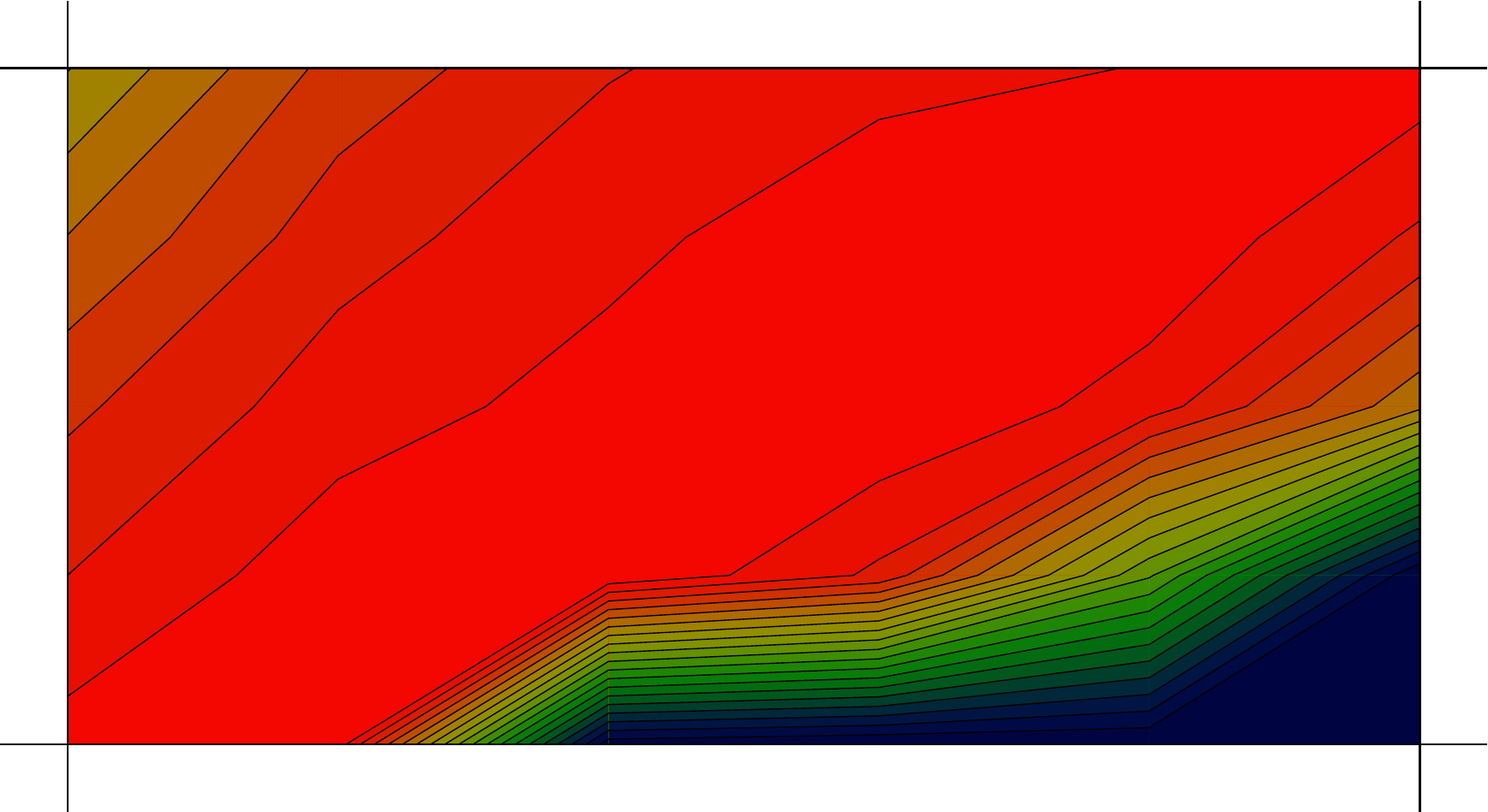}}
\put(0,50){\includegraphics[width=\hsize]{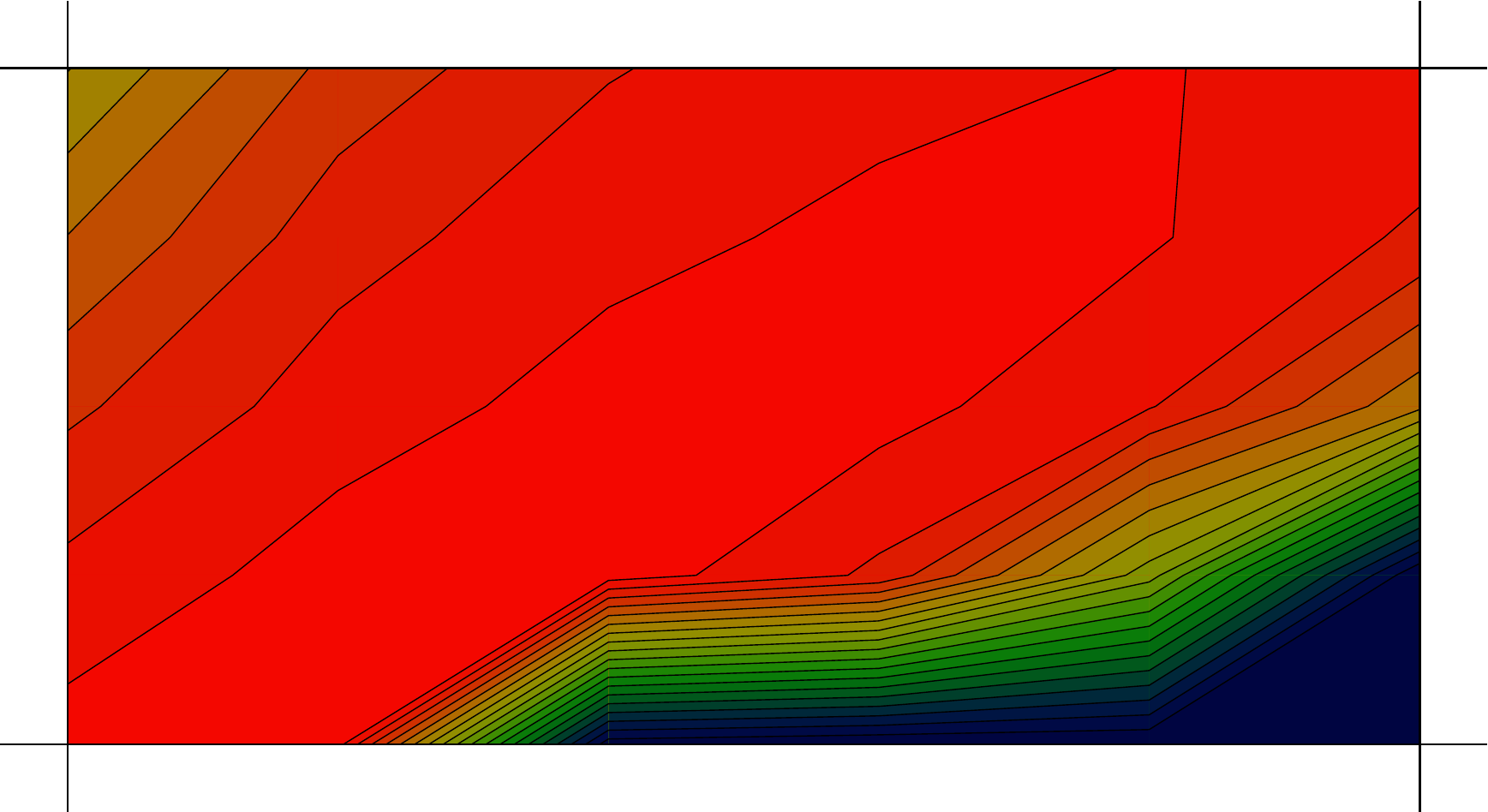}}
\put(0,100){\includegraphics[width=\hsize]{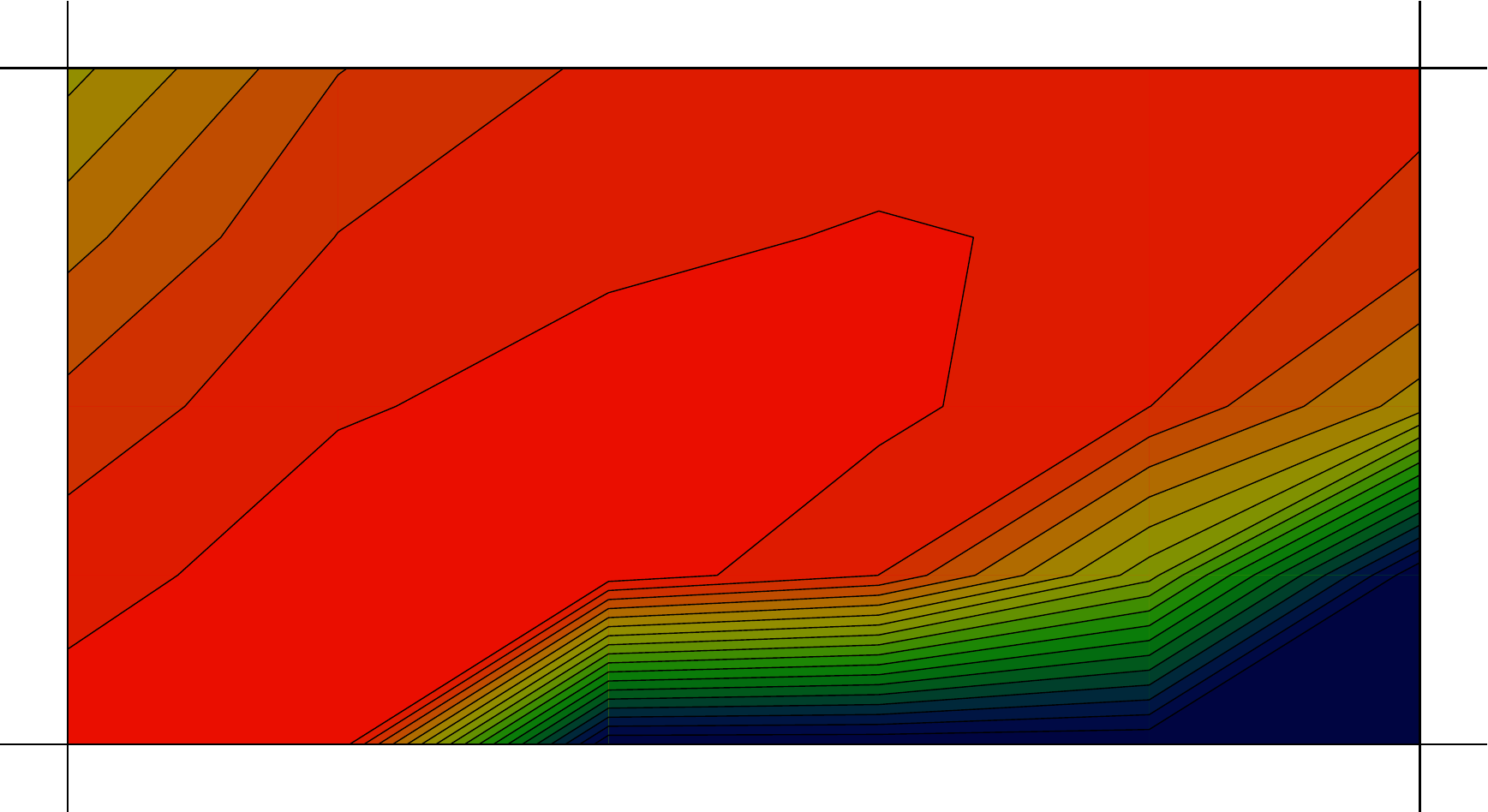}}
\put(65,0){\small $T_{\rm eff}$}
\put(0,146){\small $\log g$}
\put(4.5,1){\small 17}
\put(82,1){\small 21}
\put(1,4.5){\small 2}
\put(1,42){\small 3}
\put(1,54.5){\small 2}
\put(1,92){\small 3}
\put(1,104.5){\small 2}
\put(1,142){\small 3}
\put(12,8){\small Z=0.2}
\put(12,58){\small Z=0.5}
\put(12,108){\small Z=1.0}
\end{picture}
\caption{Residuals of the template constrained disentangling of the region 4540--4665\,\AA{} in the plane of $T_{\rm eff}$ -- $\log g$ (given in kilo-Kelvins and cgs) for different metallicities, $Z,$ of the donor star.} 
\label{Resid4540}
\end{figure}

 To determine the parameters of the atmosphere of the donor star, we disentangled the spectral region 4540--4665\,\AA{} sampled in 4096 with a step 2\,km/s/bin.
This region is dominated by the metallic lines of Si\,III, O\,II, and N\,II.
The process of disentangling into three components gives a reasonable spectrum of the donor (which \lc{is a bit less clear} but again blue-shifted and enhanced in the minima spectrum of the circumstellar component), while the gainer's spectral features are undetectable in the achieved S/N in this spectral region.
The best fit of the disentangled spectrum of the donor is shown in Fig.~\ref{Fit4540}.
This region is not especially sensitive to $T_{\rm eff}$, $\log g,$ or the metallicity $Z$, but it shows a correlation between the first two parameters (cf. Fig.~\ref{Resid4540}). 

\begin{figure}
\includegraphics[width=\hsize]{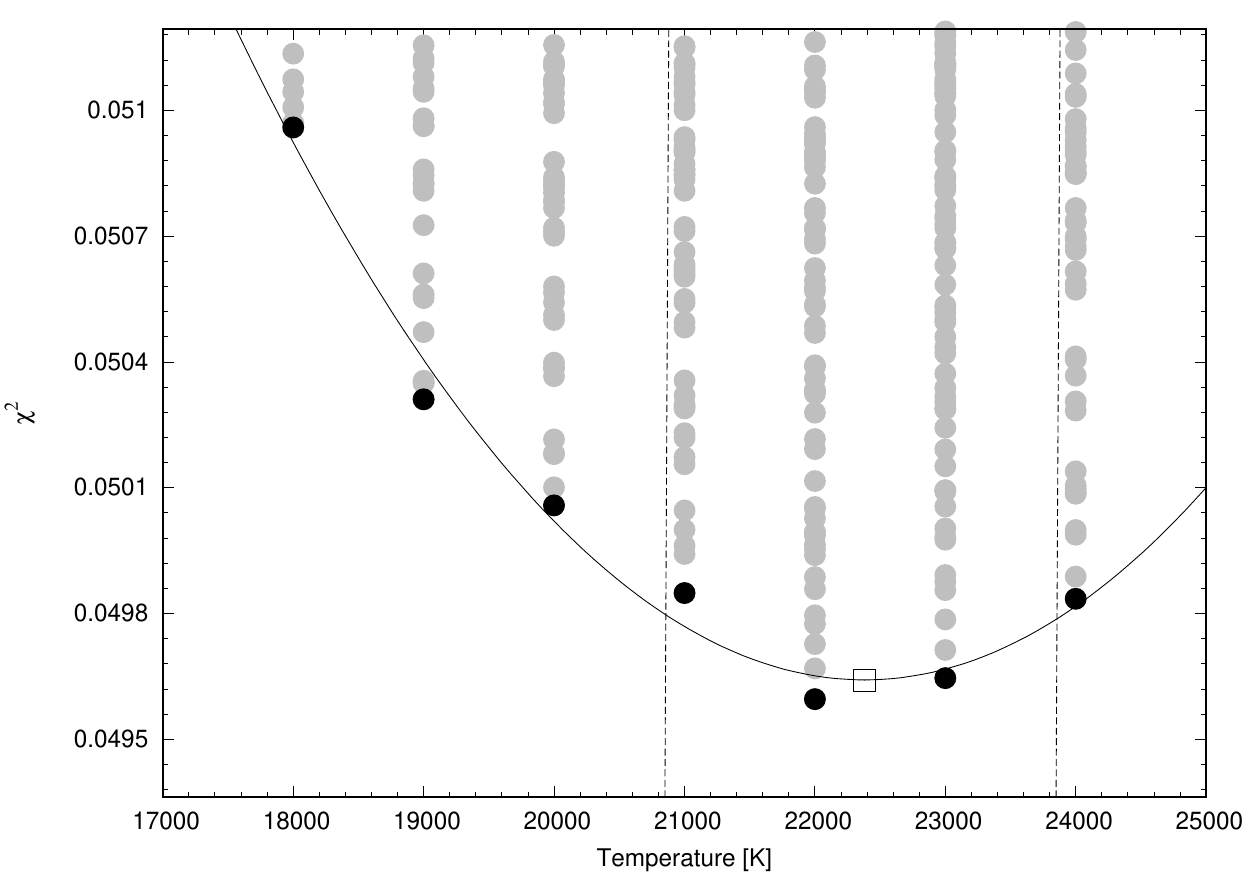}
\caption{Distribution of $\chi^2$ of the fitting for effective temperature. Black dots mark the minimal value for each temperature, gray dots show residuals for all fits, solid black line displays the quadratic fit to the infima, dashed black lines border the interval with $1\sigma$ confidence level, and the black square is the best value.}
\label{Temp}
\end{figure}

 The shorter wavelength \lc{part, approximately 4540--4585\,\AA,{}} of this spectral region leaves smaller residuals when fitted by the synthetic spectra.
An example of these residuals as a function of $T_{\rm eff}$ is shown in Fig.~\ref{Temp}.
Least-squares fits of the minima by quadratic functions yield the best solution $T_{eff}=22350\pm1500$ K, $\log g=3.17\pm0.25$ dex, and $v\sin i=101.03$\,km s$^{-1}$.

\begin{figure}
\setlength{\unitlength}{1mm}
\begin{picture}(150,90)
\put(0,0){\includegraphics[width=\hsize]{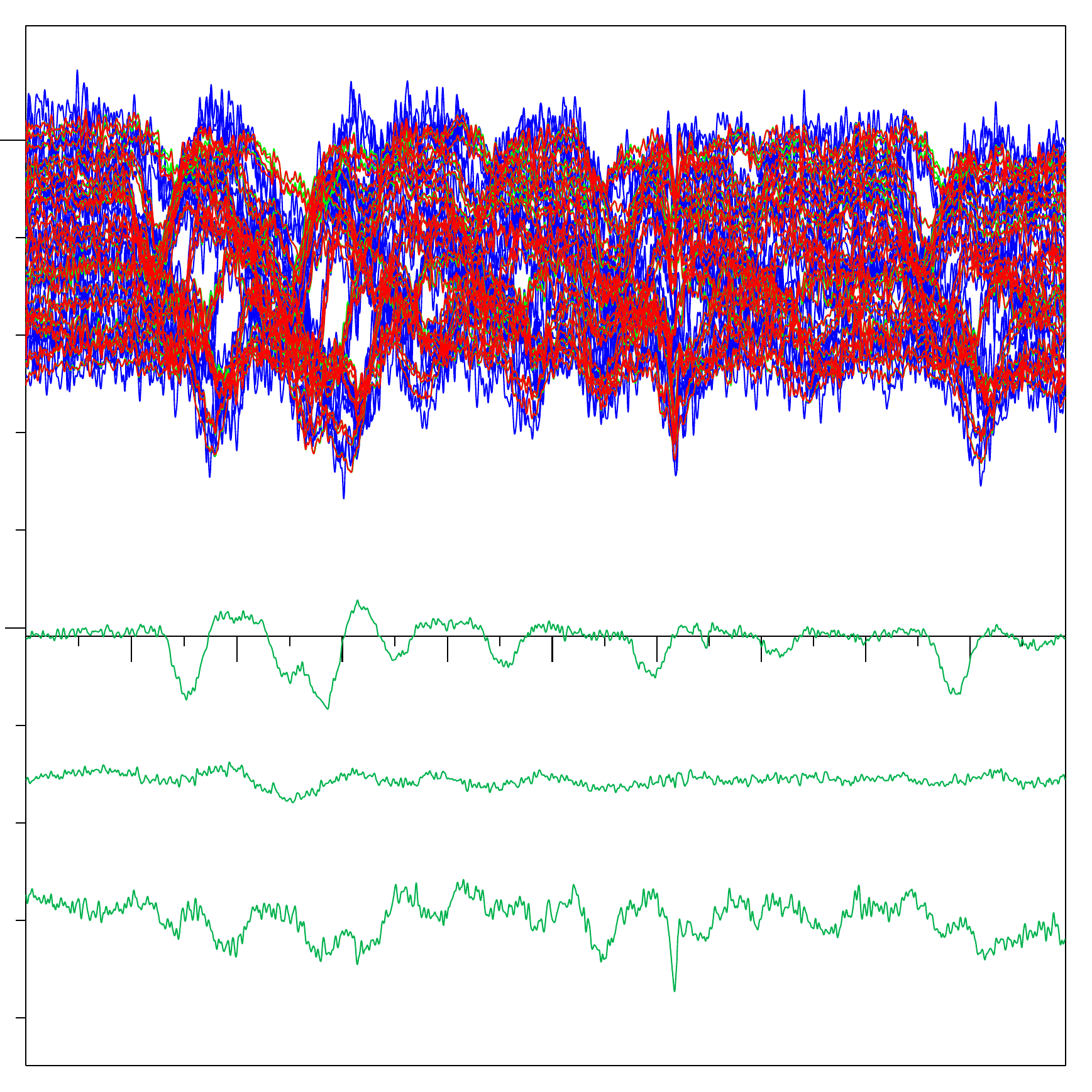}}
\put(0,82){\tiny $I$}
\put(0,75.6){\tiny 1.}
\put(0,38.5){\tiny .5}
\put(8.,38.4){\tiny 6560}
\put(42.5,38.4){\tiny 5700}
\put(77.1,38.4){\tiny 5740}
\put(82,34.5){\tiny $\lambda$}
\put(20,40){\tiny d}
\put(20,28){\tiny g}
\put(20,16){\tiny c}
\end{picture}
\caption{Disentangling of the region 5650--5749\,\AA{} in the OAN spectra of UU~Cas}           
\label{UUCas5650}
\end{figure}

 Another region we used is 5650--5749\,\AA{} (sampled in 4096 bins with a step 1.274\,km/s/bin), which is dominated by lines of N\,II, Si\,III, Al\,III, etc.
In the constrained disentangling, the lowest residual noise (0.02065 of the continuum level compared to 0.00885 for the free single-line disentangling) was found for the model with $T_{\rm eff}=23000$K, log $g=3.0$ and $Z=0.5\, Z_\odot$.
The $\gamma$-velocity of the primary has been found equal to $-66.9$\,km/s and the rotational broadening is 117.9\,km/s.
A least-squares fit of a three-dimensional (3D) quadratic form to residuals of disentangling constrained by 31 different combinations of the effective temperature, surface gravity, and the abundances yields the best solution for $T_{\rm eff}=22700$K, log $g= 2.8,$ and $Z=0.3\, Z_\odot$.
A three-component disentangling yields only very shallow and broad lines of the gainer and again blue-shifted lines of the circumstellar component (cf. Fig.~\ref{UUCas5650}).

\begin{figure}
\setlength{\unitlength}{1mm}
\begin{picture}(150,90)
\put(0,0){\includegraphics[width=\hsize]{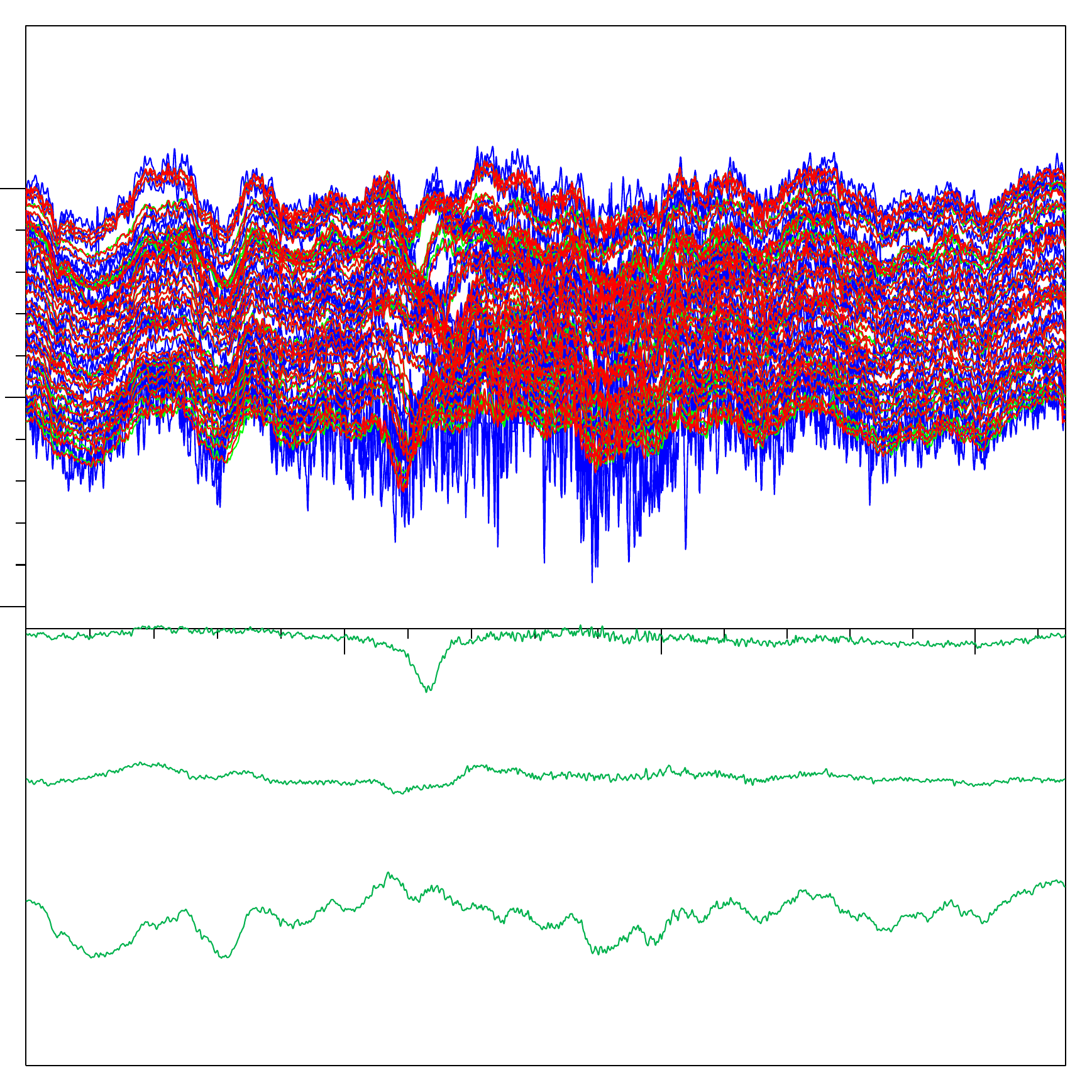}}
\put(0,82){\tiny $I$}
\put(0,75.1){\tiny 1}
\put(0,40.3){\tiny 0}
\put(25.7,38.6){\tiny 7050}
\put(51.8,38.6){\tiny 7100}
\put(77.7,38.6){\tiny 7150}
\put(82,34.5){\tiny $\lambda$}
\put(20,39){\tiny d}
\put(20,27.5){\tiny g}
\put(20,16){\tiny c}
\end{picture}
\caption{Disentangling of the region 7000--7164\,\AA{} in the OAN spectra of UU~Cas}
\label{UUCas7000}
\end{figure}

 A slightly different picture can be seen in the far-red region (cf. Fig.~\ref{UUCas7000}). 
The donor is represented in this region by He\,I line 7066\,\AA,{} which is faded at phase 0.2 and partly up to phase 0.4, while it is enhanced in both minima.
The spectrum of the gainer is again very weak, while the circumstellar component includes many strong lines.


\section{Discussion}\label{SDis}

\subsection{Disk wind}
It was previously noted by \citet{2011BlgAJ..15...87M} 
that the He\,I 5875 line is enhanced and also blue-shifted in phases around the primary minimum, where a fading would be rather expected as a consequence of the eclipse.
A strengthening of the equivalent line width of a partially eclipsed component is observed in some binaries and can be understood in terms \lc{of limb darkening} in lines \citep[cf. ][]{1997A&AS..122..581H}, 
however, the disentangling of the currently available richer set of spectra reveals that in the case of UU~Cas, we are dealing another effect altogether.

 The line strengths of the He lines of the donor are effectively diminished at the primary minimum and also at phase around 0.2, which can be due to obscuration by the gaseous stream (cf. Fig.~\ref{sf}).
The overall enhancement of the lines in the primary minimum is, however, due to a blue-shifted component.
This blue-shifted component, which can be found also in other spectral lines, has phase-locked variations in strength, but it does not follow the orbital motion of any of the components, nor any other significant phase-locked changes in the RVs.
Thus, this effect cannot be attributed to the stream from the $L_1$-point, which should be red-shifted in a projection on the gainer.
This suggests that it originates in the circumstellar matter, which is at least approximately stationary with respect to the centre of mass of the system, while the \lc{line-strength variations} in the absorption lines are due to the change of geometry of the stellar disks in its background.
In accordance with the explanation from \citet{2011BlgAJ..15...87M}, 
in terms of a velocity gradient, the blue shift indicates that this absorption originates in a stellar wind coming from either the more massive and, hence, more slowly moving gainer or, more probably, from the accretion disk around it.
Outflows of gas above and below the orbital planes have been observed also in other binaries \citep[cf. e.g.][]{1996A&A...312..879H,2012MNRAS.427..607M}.
Such disk winds flowing to both sides of the plane of an accretion disk may be driven either via magnetohydrodynamics or by radiative pressure in the lines \citep[cf. e.g.][and citations therein]{1997ApJ...477..368P,2000prpl.conf..759K,2011ApJ...736...17P,2018MNRAS.481.2628W}.

The 3D structure of the circumstellar matter violates the assumptions of Doppler tomography and mapping of the accretion disks.
A proper modelling of the dynamics and radiation of the circumstellar matter would require a 3D radiation-hydrodynamic code.
Nevertheless, it was suggested by \citet{2019ApJ...883..186K} 
that a P-Cyg profile originating in such a wind is one of the components constituting the final profile of the H$\alpha$ line.
This seems really to be \lc{the case for} the KAO spectra of 2017 (cf. Fig.~\ref{K6540}), however, the spectra taken in other seasons show a slightly different structure for H$\alpha$.

\subsection{Mass ratio}
The irregular variations in the line profile of H$\alpha$ are obviously a consequence of the changing structure of the circumstellar matter, which must influence the other spectral lines to some extent as well.
This indicates a violation of \lc{the assumptions of} the method of disentangling that splits each one from a set of observed spectra into several components with orbital motion or line profile changes known up to a few free parameters.
This violation may be different for different lines in different observational seasons.
It thus comes as no surprise that the above-listed solutions differ in some numerical results, in particular with regard to the mass ratio, which is dependent on the orbital motion of the poorly distinguishable lines of the gainer.
This difficulty limits the precision and reliability of the current results, which indicate the mass of the gainer to be about twice or more higher than the mass of the donor.
It is thus worth to check this result by an independent method.
Such a possibility is yielded also by a solution of light curves, which enables us to find the radii of the donor star and of the accretion disk.
The evidence of the presence of the circumstellar matter in the system shows that the donor component is close to its Roche lobe, the size of which depends on the mass ratio of the two stars.
This calculation given in the \lc{Appendix~\ref{AppA}} results in a value consistent with the lower limit, $Q\simeq 2,$ found from the spectroscopy.
On the other hand, the large outer radius of the accretion disk indicates that the gainer should be at least twice (or even thrice) as massive as the donor (cf. \lc{Appendix~\ref{AppA}} for a discussion of stability of the outer edge of a disk in binaries).

  To solve the remaining open question regarding the precise value of the mass ratio in UU~Cas, more spectra obtained at different states of the varying mass transfer in the system and 3D modelling of the line profile formation would be desirable.
With the data that are currently available,  we have to consider a wide range of possible masses of the component stars displayed in Fig.~\ref{sols}.
The lines denoted $K_d$ or $Q$ in this figure correspond to the semiamplitude of donor RVs or the mass ratio, respectively, fixed to the values given by the final mean solution in Table~\ref{Tab0}, and the closed curve corresponds to their 1-$\sigma$ error ellipse.
The current state of the system UU~Cas can thus be explained by a variety of initial conditions for evolution of interacting binaries.

\subsection{Evolution of the system}
In order to find the possible evolutionary state of UU~Cas, we used a 1D open source stellar evolution code named Modules for Experiments in Stellar Astrophysics \citep[MESA, ][]{2011ApJS..192....3P,2015ApJS..220...15P}, 
version r15140, compiled by MESA SDK\footnote{\url{http://www.astro.wisc.edu/~townsend/static.php?ref=mesasdk}} version 21.4.1 \citep{2020zndo...3706650T}. 
The MESA code is able to  simultaneously calculate the evolutionary path and parameters of both stars within a binary system. 
We ran around 25 different evolutionary models with different input parameters, where we mainly varied the initial masses and period. 
For all our models, we assumed a solar metallicity of $Z=0.02$.
We used the Kolb scheme \citep{1990A&A...236..385K} 
as the \lc{mass-transfer scheme} and we kept the \lc{mass-transfer efficiency} fixed by the following parameters: $\alpha=0.20$, $\beta=0.20$, $\gamma=0.05$, and $\delta=0.00$ \citep{1997A&A...327..620S}. 
Our choice is motivated by the fact that only the polar regions of the hotter gainer are \lc{open and free for} its stellar wind, while in the equatorial regions, the wind is suppressed by the accretion from the inner parts of the disk.
In any case, in the presence of disk wind, the mass loss from the system should be treated in terms of the complexity of the dynamics of the circumstellar matter, rather than a superposition of stellar winds from the component stars \citep[cf. e.g.][]{2010ApJ...708L...5M}. 
Regarding the observed speed of the wind, which is small enough in comparison with the orbital velocities of the component stars, it is obvious that the angular momentum taken away from the binary by the mass loss can be significantly influenced by the gravitational and, possibly, also the magnetic field in the system.
Thus, these current computations of the evolution of UU~Cas are only preliminary.

 The computations made using the MESA code 
of the evolution of UU~Cas indicate that the initial masses of component stars, $M_d=11.5 M_\odot$, $M_g=9.75 M_\odot$ at ZAMS, along with the initial orbital period, $P=3.5$ days, can evolve to masses close to the currently observed values $M_d=5.24 M_\odot$, $M_g=13.33 M_\odot,$ when the period increases due to the mass transfer to the present value 8.519296 days (cf. Model 1 in Table~\ref{Tabevol} and the mark 1 in Fig.~\ref{sols}).
However, several other combinations of initial conditions listed in Table~\ref{Tabevol} can also fit into the region that is compatible with the observed parameters within their uncertainty (cf. Fig.~\ref{sols}).
A linear fit to these models allows us to approximate the masses
\begin{eqnarray}
M_d(t)&\simeq&-4.24 M_\odot+0.241M_d(0)+0.390M_g(0)\nonumber\\
        &&+0.833 M_\odot P(0)/day\; ,\nonumber\\
M_g(t)&\simeq&2.33 M_\odot+0.417M_d(0)+0.786M_g(0)\nonumber\\
        &&-0.458 M_\odot P(0)/day\; .\nonumber
\end{eqnarray}

\begin{table}
\caption{Models of evolution of UU~Cas}\label{Tabevol} 
\begin{tabular}{lcccccccccc}
model& $P(0)$& $M_d(0)$& $M_g(0)$& $M_d(t)$& $M_g(t)$\\
    1&  3.5 &   11.50 &    9.75 &   5.230 &  13.199 \\ 
    2&  5.1 &   15.75 &   12.75 &   8.829 &  16.556 \\ 
    3&  4.0 &   12.00 &    9.50 &   5.679 &  12.976 \\ 
    4&  3.0 &   12.00 &    9.75 &   4.879 &  13.666 \\ 
    5&  2.8 &   10.00 &    9.10 &   4.164 &  12.310 \\ 
\end{tabular}
\tablefoot{$P(0)$ means initial orbital period in days, $M_{d,g}(0)$ initial masses and $M_{d,g}(t)$ masses in units of $M_\odot$ of the component stars when the period increases to the observed period.}
\end{table}


\section{Conclusions}\label{concl}

The present spectroscopic study confirms our previous findings based on a photometric study \citep{2020A&A...642A.211M}
on the nature of the eclipsing binary UU~Cas: 
(1) the system is in a stage of mass-exchange,
(2) it consists of two B-stars, 
(3) the brighter donor is less massive, cooler, and it fills its Roche-lobe,
(4) the heavier and hotter gainer is embedded in and partly obscured by an accretion disk.
In addition to this, our spectroscopy reveals the presence of wind, which manifests itself through a third blue-shifted absorption component in the line profiles.
The RVs of this component do not reflect the orbital motion of the binary, but the line strengths of the wind are enhanced at the phases of both \lc{light-curve minima}.
This indicates that the wind comes out from the accretion disk, in particular, from the region around the centre of mass of the system.
UU~Cas is thus in an evolutionary stage of non-conservative mass transfer and it resembles the stars of the $\beta\,$Lyr type.

 The \lc{line-profile variations}, particularly the most pronounced variations of H$\alpha$ emission, reveal that the structure of the circumstellar matter and, hence, probably the \lc{mass transfer} that feeds it as well, is not steady and, rather, it changes on a timescale of several orbital periods.
The currently available spectroscopic data do not display the long period discovered in the photometric double-periodicity of UU~Cas, but the phase coverage by spectroscopic observations is not yet sufficient to exclude it either.
We recall here that significant changes on timescales shorter than the orbital period have also been found in the photometry, which shows that the dynamics of the circumstellar matter in the system is quite complicated.


 This observational evidence confirms that UU~Cas is an interesting system that is important for studies of non-conservative mass transfer in close binary stars.
It is worthy of both long-term monitoring as well as compact observational campaigns by means of high-resolution spectroscopy.
Regarding the theoretical treatment, a 3D radiation (magneto-) hydrodynamic modelling of the \lc{mass transfer} and a computation of the \lc{synthetic spectra and light curves resulting from these models} would be highly desirable.
The \lc{mass-loss rate and angular-momentum-loss rate} obtained in this way could then help indicate better parameters for models of evolution for this and similar systems of interacting close binaries.


\begin{acknowledgements}
The authors {highly appreciate valuable comments and suggestions by the referee Herman Hensberge, which helped to improve this paper.}
We acknowledge the allotment of observing time at San Pedro M\'{a}rtir observatory, UNAM, Baja California, Mexico.
We are also indebted to Drs. Gyula Szab\'{o} and his colleagues for obtaining the GAO spectra, Zlatan Zvetanov for the APO spectra, and  Brankica Kubátová, Petr Kabáth and other colleagues from AI for the AIO spectra.
P. Hadrava and M. Cabezas were supported by project RVO 67985815.
G. Djura\v{s}evi\'{c}, J. Petrovi\'{c} and M. I. Jurkovic acknowledge the financial support from the Ministry of Education, Science and Technological Development of the Republic of Serbia through contract No. 451-03-68/2020-14/200002.
J. Garc\'{e}s acknowledges support by ANID project 21202285.
S. Yu. Gorda was supported in part by the Ministry of Science and Higher Education of the Russian Federation (projects no. FEUZ-2020-0030 and no. 075-15-2020-780, contract 780-10).
H. Markov acknowledges the support by Bulgarian National Science Fund under contract DN 18/13-12.12.2017.
R. E. Mennickent acknowledges support by BASAL Centro de Astrof{\'{i}}sica y Tecnolog{\'{i}}as Afines (CATA) PFB--06/2007, FONDECYT 1190621 and the grant ANID PIA/BASAL FB210003. 
S. Zharikov acknowledges PAPIIT grants IN102120 and Grant No. AP08856419 of the Science Committee of the Ministry of Education
and Science of the Republic of Kazakhstan.
\end{acknowledgements}

\balance
\bibliography{AAA} 
\bibliographystyle{aa} 

\begin{appendix} 
\balance
\section{Indirect methods of the \lc{mass-ratio determination}}\label{AppA}
\begin{figure}
\includegraphics[width=\hsize]{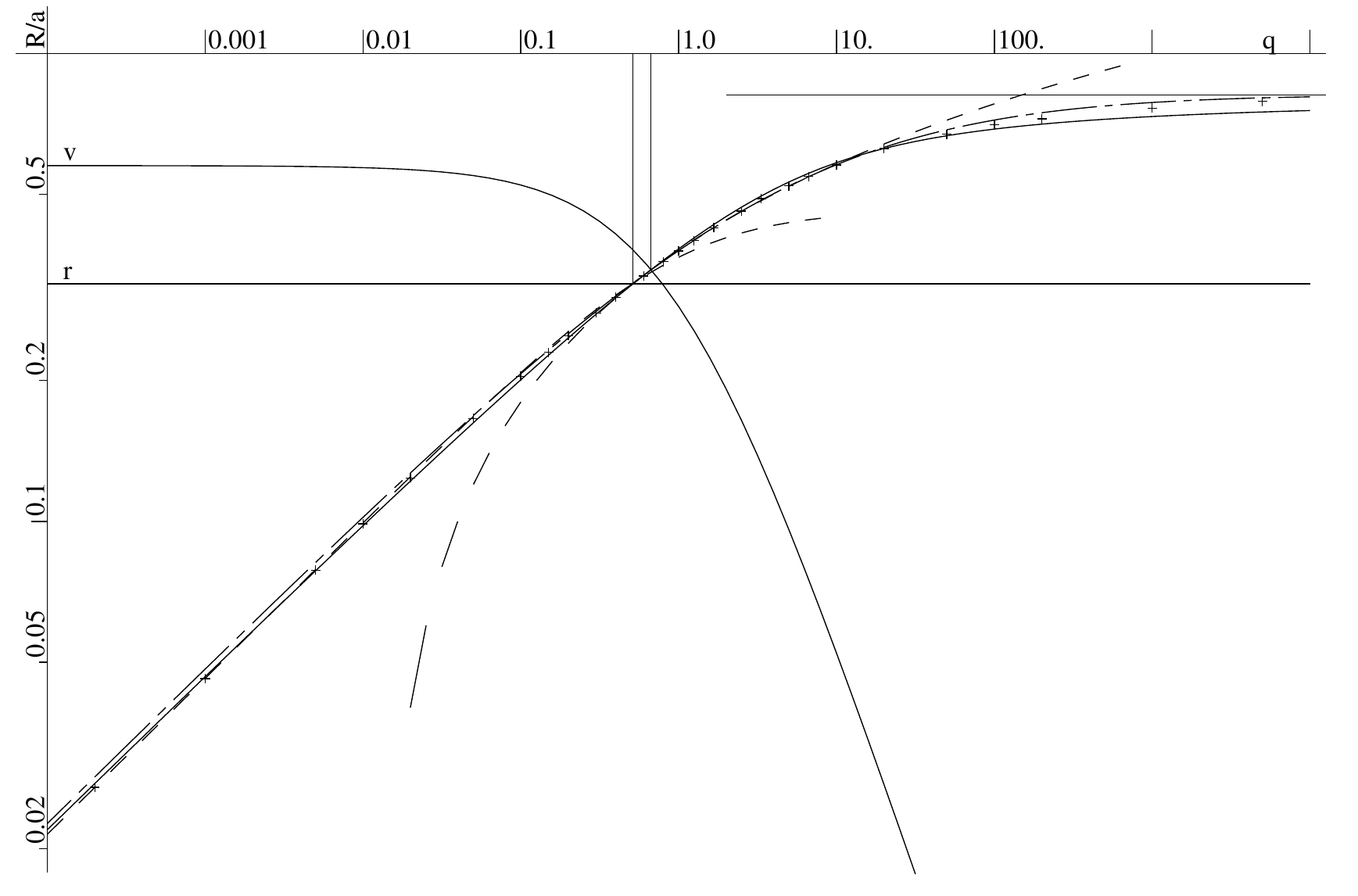}
\caption{Relative radius of Roche lobe in logarithmic scale as a function of the mass ratio 
according to calculations by Kopal (+) and their fits by Paczy\'{n}ski (dashed lines), 
Hadrava (full line), and Eggleton (dash-dotted line).
Solution for UU~Cas according to photometric radius (line r) and rotational broadening
(curve v)}
\label{UUCasq}  
\end{figure}

The straightforward determination of the mass ratio from amplitudes of RV-curves of both components can be verified using methods of an indirect estimate.
If we assume that the donor star fills its Roche lobe, then the ratio of its radius $R_d$ to the distance $a$ of centres of the component stars depends on the mass ratio $q\equiv M_d/M_g=1/Q$ of the two stars.
If we find $R_d/a$ from photometry, we can thus infer that $q$ should not be higher than the value at which the Roche lobe would constrict the donor.
The volume of the Roche lobe and radius of a sphere of the same volume as a function of $q$ has been calculated and tabulated by \citet[p. 136]{1959cbs..book.....K}. 
Several formulae have been introduced \citep{1971ARA&A...9..183P,1980PhDT.........0H,1984BAICz..35..335H,1983ApJ...268..368E} 
to approximate this dependence with a precision higher than the sphere approximates the Roche lobe (cf. Fig.~\ref{UUCasq}).
The formula by Hadrava 
\begin{equation}
 \log\frac{R_d}{a}=\frac{\log q}{6}-c_1-\left[\left(\frac{\log q}{6}-c_2\right)^2+c_3\right]^\frac{1}{2}\; ,
\end{equation}
where $c_1=0.183148$, $c_2=0.110097,$ and $c_3=0.04412135$ are slightly improved values of parameters, can be analytically solved for $q$ in the form
\begin{equation}\label{Rochinv}
 \log q= 3 \frac{(c_1+\log \frac{R_d}{a})^2-c_4}{c_5+\log\frac{R_d}{a}}\; ,
\end{equation}
where $c_4=c_2^2+c_3=0.0562427$ and $c_5=c_1-c_2=0.073051$.
We note that to get the correct assymptotic radius $R_d/a=(3\sqrt{3}-4+3\ln(6-3\sqrt{3}))^{1/3}$ for $q\rightarrow\infty$ (shown by a horizontal line in Fig.~\ref{UUCasq}), $c_5$ should be fixed to its negative logarithm which is $\sim 0.0889033$. 
This would, however, result in a less precise fit at $q$ around 1, which \lc{is more needed in} calculations for binary stars.

 The photometric solution \citep[Table 2]{2020A&A...642A.211M} gives, for the case of UU~Cas:
$R_d=16.9 R_\odot$ and $a=52.2 R_\odot$, that is, $R_d/a=0.324$. From the solution (\ref{Rochinv}), 
we then find $q=M_d/M_g=0.535$, {that is, $Q=M_g/M_d=1.87$,} which is marked in Fig.~\ref{UUCasq} 
by the line denoted as `r'. 

 There is also an alternative possibility to estimate the radius of a Roche-lobe
filling star from the rotational broadening $V_{rot}$ of its spectral lines.
If the star corotates with the orbit, as it is the standard assumption of the Roche model, 
then $R_d/a_d=V_{rot}/K_d$, where $K_d$ is the semiamplitude of the RV-curve 
of the donor and $a_d=a/(1+q)$ is radius of its orbit around the centre of mass.
The intersection of the curve $R_d/a=V_{rot}/(K_d(1+q))$ denoted by `v' in Fig.~\ref{UUCasq}, 
with the curve of the relative radius of the Roche lobe, defines this spectroscopic mass
ratio. By disentangling  the line He\,I 5875 for UU~Cas, we found $V_{rot}/K_d=0.5756$, 
which results in the {mass ratio $Q=M_g/M_d=1.496$.}
However, in the disentangled line profile of the donor, we can see weak emission 
wings separated for about 170\,km/s from the centre of the line.
This may indicate an extended rotating envelope of the donor star and its non-synchronous 
rotation, which would invalidate this spectroscopic determination of $q$.
A non-synchronous rotation of the donor would play a role in the dynamics of the circumstellar 
matter and complicate the interpretation of both the spectroscopy as well as photometry.

 Yet another constraint on the mass ratio can be obtained from the size of the accretion disk
around the gainer.
According to 
\citet[Table 2]{2020A&A...642A.211M}, the outer radius of the disk is 20.8\,$R_\odot=0.4\,a$.
The Roche lobe of the gainer thus must be larger, which is satisfied if $M_g>1.2\,M_d$.\vspace*{0.2mm}

 The outer edge of the disk, however, cannot reach the Roche lobe because of perturbations of streamlines in its outer parts.
This implies a more stringent limitation on the mass ratio.
Assuming \lc{that the streamlines follow} the simple periodic orbits, \citet{1977ApJ...216..822P} 
imposed a criterion on the maximum size of the disk to be given by the largest orbit that does not intersect its neighbouring orbits.
These simple periodic orbits are not circular.
However, if we approximate their radii by the mean values between their minima and maxima, we find the disk size to be 0.4 of the distance between the component stars to correspond to $M_g\simeq 2.6\,M_d$.
It means that $Q$ should be 2.6 or more to enable (according to Paczy{\'n}ski's criterion) the existence of a stable disk
of a size of $0.4a$.
On the other hand, the spectroscopic evidence for the \lc{disk wind} may indicate that the outer rim of the photometrically detected disk is unstable due to a slight violation of this criterion.

 For mass ratios of $Q>4.5$, the alternative criterion of dynamical stability of the simple periodic orbits restricts the size of the accretion disk \lc{more} than the Paczy{\'n}ski's criterion \citep[to about 0.46\,$a$ {at maximum}, cf.][]{1990CeMDA..48..115K}. 
Hence, the dynamical instability would not allow much larger disk, even for a higher mass of the gainer, which would thus also be consistent with the observed size of the accretion disk in UU~Cas.
\clearpage

\onecolumn

\section{Additional table}

\begin{longtable}{rrrrrrrr}
\caption{\label{Tab1}Radial velocities and line-strength factors from the He\,I lines 5875, 6678, and 7065\,{\AA} in OAN, KAO, and AIO spectra (according to Solutions 8, 11, and 12)}\\
\hline
$n$&\multicolumn{1}{c}{$T_{exp}$}& phase& \multicolumn{1}{c}{$RV_d$}& \multicolumn{1}{c}{$RV_g$}& \multicolumn{1}{c}{$\exp(s_1)$}& \multicolumn{1}{c}{$\exp(s_2)$}& \multicolumn{1}{c}{$\exp(s_3)$}\\
 \hline
\endfirsthead
\caption{continued}\\
$n$&\multicolumn{1}{c}{$T_{exp}$}& phase& \multicolumn{1}{c}{$RV_d$}& \multicolumn{1}{c}{$RV_g$}& \multicolumn{1}{c}{$\exp(s_1)$}& \multicolumn{1}{c}{$\exp(s_2)$}& \multicolumn{1}{c}{$\exp(s_3)$}\\
\hline
\endhead
\endfoot
\multicolumn{8}{c}{OAN}\\
  1& 5580.6136& 0.184& $-169.6\pn 6.9$& $ 48.3\pn 1.4$& $0.16\pm0.05$& $0.50\pm0.16$& $0.44\pm0.02$\\ 
  2& 5580.6266& 0.185& $-171.5\pn 8.4$& $ 49.0\pm16.6$& $0.20\pm0.17$& $0.34\pm0.13$& $0.36\pm0.13$\\ 
  3& 5580.6395& 0.187& $-172.8\pn 4.9$& $ 53.4\pm10.6$& $0.16\pm0.08$& $0.41\pm0.17$& $0.37\pm0.05$\\ 
  4& 5581.6386& 0.304& $-170.2\pn 2.2$& $ 70.1\pm33.1$& $0.52\pm0.23$& $0.33\pm0.30$& $0.10\pm0.07$\\ 
  5& 5581.6518& 0.305& $-172.0\pn 3.6$& $ 73.3\pm40.6$& $0.51\pm0.22$& $0.30\pm0.25$& $0.07\pm0.06$\\ 
  6& 5581.6651& 0.307& $-170.2\pn 8.8$& $ 67.6\pm20.9$& $0.47\pm0.25$& $0.27\pm0.27$& $0.07\pm0.09$\\ 
  7& 5582.6165& 0.418& $ -87.3\pn 0.9$& $ 26.0\pn 1.6$& $0.80\pm0.14$& $0.18\pm0.19$& $0.05\pm0.07$\\ 
  8& 5582.6298& 0.419& $ -87.2\pn 2.1$& $ 25.6\pn 6.6$& $0.81\pm0.16$& $0.22\pm0.25$& $0.05\pm0.07$\\ 
  9& 5582.6428& 0.421& $ -86.2\pn 1.9$& $ 22.5\pm14.4$& $0.81\pm0.15$& $0.21\pm0.14$& $0.04\pm0.05$\\ 
 10& 5583.5988& 0.533& $  24.9\pm17.0$& $ -1.6\pm23.1$& $0.72\pm0.11$& $0.36\pm0.17$& $0.30\pm0.12$\\ 
 11& 5583.6140& 0.535& $  30.7\pm11.3$& $-14.8\pn 2.4$& $0.60\pm0.19$& $0.63\pm0.29$& $0.31\pm0.08$\\ 
 12& 5583.6273& 0.536& $  36.6\pn 8.6$& $-19.1\pn 4.9$& $0.73\pm0.12$& $0.44\pm0.15$& $0.37\pm0.06$\\ 
 13& 5584.5968& 0.650& $ 159.0\pn 6.1$& $-56.2\pn 6.6$& $0.59\pm0.04$& $0.49\pm0.02$& $0.23\pm0.08$\\ 
 14& 5584.6098& 0.651& $ 156.7\pn 4.5$& $-57.2\pn 7.8$& $0.60\pm0.06$& $0.46\pm0.04$& $0.28\pm0.06$\\ 
 15& 5584.6227& 0.653& $ 162.7\pn 4.2$& $-60.2\pm11.1$& $0.56\pm0.03$& $0.47\pm0.12$& $0.22\pm0.07$\\ 
 16& 5585.6282& 0.770& $ 197.9\pn 4.0$& $-68.6\pm10.1$& $0.64\pm0.04$& $0.24\pm0.17$& $0.15\pm0.05$\\ 
 17& 5585.6411& 0.772& $ 193.5\pm10.6$& $-70.1\pm13.8$& $0.61\pm0.04$& $0.19\pm0.03$& $0.10\pm0.06$\\ 
 18& 5585.6541& 0.773& $ 197.5\pn 5.0$& $-65.8\pm11.4$& $0.69\pm0.03$& $0.23\pm0.05$& $0.10\pm0.08$\\ 
 19& 5586.6031& 0.884& $ 122.7\pn 4.0$& $-43.2\pn 3.8$& $0.76\pm0.11$& $0.35\pm0.12$& $0.11\pm0.07$\\ 
 20& 5586.6148& 0.886& $ 119.8\pn 1.5$& $-39.5\pn 9.2$& $0.78\pm0.08$& $0.29\pm0.12$& $0.18\pm0.02$\\ 
 21& 5586.6266& 0.887& $ 119.9\pn 4.4$& $-38.4\pn 7.5$& $0.76\pm0.08$& $0.32\pm0.17$& $0.11\pm0.07$\\ 
 22& 5587.5959& 0.000& $  -8.4\pm13.9$& $-29.1\pn 1.2$& $0.47\pm0.11$& $0.13\pm0.18$& $0.67\pm0.14$\\ 
 23& 5587.6077& 0.002& $ -11.0\pm10.4$& $  2.4\pn 0.3$& $0.47\pm0.11$& $0.12\pm0.15$& $0.68\pm0.15$\\ 
 24& 5587.6194& 0.003& $ -11.4\pm10.1$& $ -4.4\pn 2.3$& $0.42\pm0.12$& $0.16\pm0.21$& $0.73\pm0.11$\\ 
 25& 5588.5988& 0.118& $-132.7\pn 4.8$& $ 50.4\pn 8.7$& $0.54\pm0.23$& $0.53\pm0.20$& $0.36\pm0.04$\\ 
 26& 5588.6105& 0.119& $-132.2\pn 6.3$& $ 58.3\pm14.8$& $0.56\pm0.22$& $0.44\pm0.16$& $0.33\pm0.04$\\ 
 27& 5588.6222& 0.120& $-132.3\pn 4.3$& $ 50.2\pn 5.5$& $0.58\pm0.27$& $0.47\pm0.11$& $0.32\pm0.05$\\ 
\hline 
\multicolumn{8}{c}{KAO}\\
  1& 7822.1542& 0.309& $-183.8\pn 3.0$& $ 68.3\pn 8.6$& $0.56\pm0.13$& $0.19\pm0.03$& $0.05\pm0.06$\\
  2& 7827.1868& 0.900& $ 122.3\pn 1.4$& $-43.0\pn 3.4$& $0.84\pm0.09$& $0.38\pm0.16$& $0.22\pm0.02$\\
  3& 7827.5054& 0.937& $  86.9\pn 2.9$& $-30.8\pn 1.0$& $0.81\pm0.12$& $0.38\pm0.23$& $0.25\pm0.10$\\
  4& 7828.1918& 0.018& $  -6.3\pn 3.8$& $  3.7\pn 0.2$& $0.47\pm0.08$& $0.02\pm0.03$& $0.77\pm0.05$\\
  5& 7828.5054& 0.054& $ -61.2\pm17.0$& $-18.1\pm39.4$& $0.19\pm0.07$& $0.06\pm0.04$& $0.84\pm0.10$\\
  6& 7829.1791& 0.133& $-141.9\pn 4.6$& $ 46.0\pn 4.4$& $0.53\pm0.26$& $0.30\pm0.05$& $0.26\pm0.16$\\
  7& 7829.4954& 0.171& $-160.2\pn 4.9$& $ 58.4\pn 6.6$& $0.41\pm0.18$& $0.37\pm0.08$& $0.28\pm0.09$\\
  8& 7832.1714& 0.485& $ -31.0\pn 3.5$& $ 10.8\pn 1.2$& $0.69\pm0.07$& $0.32\pm0.16$& $0.14\pm0.04$\\
  9& 7832.4728& 0.520& $  12.1\pn 4.7$& $ -3.7\pn 1.6$& $0.72\pm0.04$& $0.29\pm0.07$& $0.19\pm0.03$\\
 10& 7834.1744& 0.720& $ 185.1\pn 4.1$& $-65.1\pn 4.9$& $0.57\pm0.05$& $0.47\pm0.17$& $0.27\pm0.07$\\
 11& 7834.5053& 0.759& $ 191.7\pn 2.8$& $-66.3\pn 4.3$& $0.61\pm0.04$& $0.41\pm0.15$& $0.25\pm0.01$\\
 12& 7835.1758& 0.837& $ 166.0\pn 4.2$& $-58.4\pn 4.9$& $0.70\pm0.07$& $0.32\pm0.10$& $0.13\pm0.02$\\
 13& 7835.4776& 0.873& $ 138.9\pn 1.3$& $-49.6\pn 6.4$& $0.81\pm0.05$& $0.30\pm0.14$& $0.15\pm0.03$\\
 14& 7870.4073& 0.973& $  45.6\pm14.3$& $-16.3\pn 2.0$& $0.40\pm0.15$& $0.74\pm0.24$& $0.27\pm0.09$\\
 15& 7962.3187& 0.761& $ 198.7\pn 9.0$& $-68.0\pn 5.2$& $0.76\pm0.09$& $0.31\pm0.22$& $0.14\pm0.01$\\
 16& 8037.1537& 0.546& $  41.0\pn 5.3$& $-19.4\pn 4.9$& $0.88\pm0.10$& $0.44\pm0.20$& $0.29\pm0.05$\\
 17& 8038.1648& 0.664& $ 149.8\pn 9.2$& $-52.2\pn 4.8$& $0.72\pm0.10$& $0.72\pm0.18$& $0.20\pm0.08$\\ \hline
 18& 8124.2656& 0.771& $ 212.2\pn 3.5$& $-70.4\pn 7.5$& $0.62\pm0.05$& $0.20\pm0.10$& $0.13\pm0.00$\\
 19& 8125.2302& 0.884& $ 151.9\pn 2.9$& $-45.5\pm20.7$& $0.78\pm0.10$& $0.38\pm0.16$& $0.17\pm0.05$\\
 20& 8131.1662& 0.581& $ 109.2\pn 3.1$& $-46.3\pm22.5$& $0.74\pm0.09$& $0.39\pm0.13$& $0.33\pm0.05$\\
 21& 8158.2304& 0.758& $ 225.4\pn 1.7$& $-66.2\pm10.8$& $0.57\pm0.06$& $0.20\pm0.13$& $0.15\pm0.05$\\
 22& 8159.1402& 0.864& $ 166.5\pn 2.5$& $-56.7\pm17.3$& $0.86\pm0.10$& $0.32\pm0.17$& $0.17\pm0.04$\\
 23& 8166.1600& 0.688& $ 164.8\pm12.7$& $-74.7\pn 9.3$& $0.54\pm0.06$& $0.60\pm0.19$& $0.08\pm0.05$\\
 24& 8197.1712& 0.329& $-133.0\pn 5.9$& $ 95.1\pn 2.3$& $0.26\pm0.08$& $0.07\pm0.06$& $0.33\pm0.02$\\
\hline
\multicolumn{8}{c}{AIO}\\
  1& 8925.5886& 0.831& $ 160.8\pn 3.2$& $-58.2\pn 7.3$& $0.88\pm0.09$& $0.17\pm0.09$& $0.19\pm0.16$\\
  2& 8930.6473& 0.424& $-109.2\pm12.4$& $ 37.1\pn 0.2$& $0.31\pm0.10$& $0.30\pm0.42$& $0.27\pm0.08$\\
  3& 8931.6326& 0.540& $  34.5\pn 4.6$& $-12.9\pn 1.9$& $0.51\pm0.11$& $0.45\pm0.25$& $0.25\pm0.18$\\
  4& 8932.6484& 0.659& $ 147.4\pn 6.0$& $-53.6\pn 4.0$& $0.67\pm0.08$& $0.49\pm0.21$& $0.17\pm0.04$\\
  5& 8933.6189& 0.773& $ 180.5\pn 7.3$& $-61.3\pn 6.4$& $0.56\pm0.03$& $0.13\pm0.03$& $0.11\pm0.04$\\
  6& 8941.6255& 0.713& $ 168.1\pn 7.7$& $-64.4\pn 4.3$& $0.56\pm0.08$& $0.52\pm0.32$& $0.18\pm0.01$\\
  7& 8945.5756& 0.177& $-172.4\pn 4.8$& $ 56.7\pn 4.6$& $0.44\pm0.13$& $0.56\pm0.19$& $0.29\pm0.11$\\
  8& 9012.4991& 0.032& $ -17.9\pn 9.7$& $  9.4\pn 1.9$& $0.40\pm0.08$& $0.21\pm0.09$& $0.99\pm0.01$\\
  9& 9013.5054& 0.150& $-154.3\pn 1.4$& $ 51.0\pn 4.3$& $0.64\pm0.43$& $0.58\pm0.16$& $0.54\pm0.33$\\
 10& 9016.4856& 0.500& $ -15.5\pn 4.9$& $  5.5\pn 3.4$& $0.67\pm0.08$& $0.11\pm0.08$& $0.46\pm0.04$\\
 11& 9024.5357& 0.445& $ -77.8\pn 2.6$& $ 24.2\pn 0.4$& $0.49\pm0.03$& $0.02\pm0.02$& $0.35\pm0.12$\\
 12& 9030.4518& 0.140& $-157.3\pn 9.4$& $ 46.6\pn 4.3$& $0.41\pm0.27$& $0.37\pm0.10$& $0.44\pm0.22$\\
 13& 9111.3418& 0.634& $ 129.2\pn 5.7$& $-46.9\pn 3.8$& $0.65\pm0.11$& $0.64\pm0.25$& $0.24\pm0.04$\\
 14& 9265.2553& 0.701& $ 185.1\pn 2.8$& $-63.0\pn 5.0$& $0.49\pm0.03$& $0.60\pm0.29$& $0.18\pm0.03$\\
 15& 9271.2486& 0.404& $-126.4\pn 9.4$& $ 41.9\pm25.7$& $0.46\pm0.11$& $0.01\pm0.02$& $0.17\pm0.06$\\
 16& 9279.2554& 0.344& $-164.7\pn 2.4$& $ 54.4\pn 6.1$& $0.30\pm0.13$& $0.11\pm0.08$& $0.18\pm0.13$\\
 17& 9294.2853& 0.108& $-147.9\pm15.0$& $106.7\pm23.3$& $0.13\pm0.09$& $0.02\pm0.03$& $0.24\pm0.07$\\
\hline 
\end{longtable}                                                                             

\end{appendix}

\end{document}